%% file: bright.tex
\documentclass[twocolumn]{aastex631}
\usepackage{xfrac,graphicx,color,float,hyperref,lineno}  
 
\newcommand {\lya}{Ly$\alpha$}
\newcommand {\oiii}{[O{\sc iii}]}
\def\ltsima{$\; \buildrel < \over \sim \;$}
\def\simlt{\lower.5ex\hbox{\ltsima}}
\def\gtsima{$\; \buildrel > \over \sim \;$}
\def\simgt{\lower.5ex\hbox{\gtsima}}
\newcommand{\gala}{COS-z10-1}
\newcommand{\galb}{COS-z12-1}
\newcommand{\galc}{COS-z12-2}
\newcommand{\gald}{COS-z12-3}
\newcommand{\gale}{COS-z10-3}
\newcommand{\galf}{COS-z10-4}
\newcommand{\galg}{COS-z11-2}
\newcommand{\galh}{COS-z11-1}
\newcommand{\gali}{COS-z14-2}
\newcommand{\galj}{COS-z13-1}
\newcommand{\galk}{COS-z13-3}
\newcommand{\gall}{COS-z13-2}
\newcommand{\galm}{COS-z14-1}
\newcommand{\galn}{COS-z10-2}
\newcommand{\galz}{COS-z12-4}

\newcommand {\etal}{et~al.}

\newcommand {\um}{$\mu$m}
\newcommand {\muv}{$M_{\rm UV}$}

\newcommand{\msun}{{\rm\,M$_\odot$}}
\newcommand{\sfr}{{\rm\,M$_\odot$\,yr$^{-1}$}}
\newcommand{\lsun}{{\rm\,L$_\odot$}}

\shorttitle{COSMOS-Web Bright $z>10$ Candidates}
\shortauthors{Casey et al.}

\begin{document}

\title{COSMOS-Web: Intrinsically Luminous $z\simgt10$ Galaxy Candidates Test Early Stellar Mass Assembly}

\correspondingauthor{Caitlin M. Casey}
\email{cmcasey@utexas.edu}

\author[0000-0002-0930-6466]{Caitlin M. Casey}
\affiliation{The University of Texas at Austin, 2515 Speedway Blvd Stop C1400, Austin, TX 78712, USA}
\affiliation{Cosmic Dawn Center (DAWN), Denmark}

\author[0000-0003-3596-8794]{Hollis B. Akins}
\affiliation{The University of Texas at Austin, 2515 Speedway Blvd Stop C1400, Austin, TX 78712, USA}

\author[0000-0002-7087-0701]{Marko Shuntov}
\affiliation{Cosmic Dawn Center (DAWN), Denmark} 
\affiliation{Niels Bohr Institute, University of Copenhagen, Jagtvej 128, DK-2200, Copenhagen, Denmark}

\author[0000-0002-7303-4397]{Olivier Ilbert}
\affiliation{Aix Marseille Universit\'{e}, CNRS, CNES, LAM, Marseille, France}

\author[0000-0003-2397-0360]{Louise Paquereau} 
\affiliation{Institut d’Astrophysique de Paris, UMR 7095, CNRS, and Sorbonne Universit\'{e}, 98 bis boulevard Arago, F-75014 Paris, France}

\author[0000-0002-3560-8599]{Maximilien Franco}
\affiliation{The University of Texas at Austin, 2515 Speedway Blvd Stop C1400, Austin, TX 78712, USA}

\author[0000-0003-4073-3236]{Christopher C. Hayward}
\affiliation{Center for Computational Astrophysics, Flatiron Institute, 162 Fifth Avenue, New York, NY 10010, USA}

\author[0000-0001-8519-1130]{Steven L. Finkelstein}
\affiliation{The University of Texas at Austin, 2515 Speedway Blvd Stop C1400, Austin, TX 78712, USA}

\author[0000-0002-9604-343X]{Michael Boylan-Kolchin}
\affiliation{The University of Texas at Austin, 2515 Speedway Blvd Stop C1400, Austin, TX 78712, USA}

\author[0000-0002-4271-0364]{Brant E. Robertson}
\affiliation{Department of Astronomy and Astrophysics, University of California, Santa Cruz, 1156 High Street, Santa Cruz, CA 95064, USA}

\author[0000-0001-9610-7950]{Natalie Allen}
\affiliation{Cosmic Dawn Center (DAWN), Denmark} 
\affiliation{Niels Bohr Institute, University of Copenhagen, Jagtvej 128, DK-2200, Copenhagen, Denmark}

\author[0000-0002-0245-6365]{Malte Brinch}
\affiliation{Cosmic Dawn Center (DAWN), Denmark} 
\affiliation{DTU-Space, National Space Institute, Technical University of Denmark, Elektrovej 327, 2800 Kgs, Lyngby, Denmark}

\author[0000-0003-3881-1397]{Olivia R. Cooper}\altaffiliation{NSF Graduate Research Fellow}
\affiliation{The University of Texas at Austin, 2515 Speedway Blvd Stop C1400, Austin, TX 78712, USA}

\author[0000-0002-0786-7307]{Xuheng Ding}
\affiliation{Kavli Institute for the Physics and Mathematics of the Universe (WPI), The University of Tokyo, Kashiwa, Chiba 277-8583, Japan}

\author[0000-0003-4761-2197]{Nicole E. Drakos}
\affiliation{Department of Astronomy and Astrophysics, University of California, Santa Cruz, 1156 High Street, Santa Cruz, CA 95064, USA}

\author[0000-0002-9382-9832]{Andreas L. Faisst}
\affiliation{Caltech/IPAC, MS 314-6, 1200 E. California Blvd. Pasadena, CA 91125, USA}

\author[0000-0001-7201-5066]{Seiji Fujimoto}\altaffiliation{NASA Hubble Fellow}
\affiliation{The University of Texas at Austin, 2515 Speedway Blvd Stop C1400, Austin, TX 78712, USA}

\author[0000-0001-9885-4589]{Steven Gillman}
\affiliation{Cosmic Dawn Center (DAWN), Denmark}
\affiliation{DTU-Space, National Space Institute, Technical University of Denmark, Elektrovej 327, 2800 Kgs, Lyngby, Denmark}

\author[0000-0003-0129-2079]{Santosh Harish}
\affiliation{Laboratory for Multiwavelength Astrophysics, School of Physics and Astronomy, Rochester Institute of Technology, 84 Lomb Memorial Drive, Rochester, NY 14623, USA}

\author[0000-0002-3301-3321]{Michaela Hirschmann}
\affiliation{Institute of Physics, GalSpec, Ecole Polytechnique Federale de Lausanne, Observatoire de Sauverny, Chemin Pegasi 51, 1290 Versoix, Switzerland}
\affiliation{INAF, Astronomical Observatory of Trieste, Via Tiepolo 11, 34131 Trieste, Italy}

\author[0000-0002-8412-7951]{Shuowen Jin}
\affiliation{Cosmic Dawn Center (DAWN), Denmark} 
\affiliation{DTU-Space, National Space Institute, Technical University of Denmark, Elektrovej 327, 2800 Kgs, Lyngby, Denmark}

\author[0000-0001-9187-3605]{Jeyhan S. Kartaltepe}
\affiliation{Laboratory for Multiwavelength Astrophysics, School of Physics and Astronomy, Rochester Institute of Technology, 84 Lomb Memorial Drive, Rochester, NY 14623, USA}

\author[0000-0002-6610-2048]{Anton M. Koekemoer}
\affiliation{Space Telescope Science Institute, 3700 San Martin Dr., Baltimore, MD 21218, USA} 

\author[0000-0002-5588-9156]{Vasily Kokorev}
\affiliation{Kapteyn Astronomical Institute, University of Groningen, PO Box 800, 9700 AV Groningen, The Netherlands}

\author[0000-0001-9773-7479]{Daizhong Liu}
\affiliation{Max-Planck-Institut f\"ur Extraterrestrische Physik (MPE), Giessenbachstr. 1, D-85748 Garching, Germany}

\author[0000-0002-7530-8857]{Arianna S. Long}\altaffiliation{NASA Hubble Fellow}
\affiliation{The University of Texas at Austin, 2515 Speedway Blvd Stop C1400, Austin, TX 78712, USA}

\author[0000-0002-4872-2294]{Georgios Magdis}
\affiliation{Cosmic Dawn Center (DAWN), Denmark} 
\affiliation{DTU-Space, National Space Institute, Technical University of Denmark, Elektrovej 327, 2800 Kgs, Lyngby, Denmark}
\affiliation{Niels Bohr Institute, University of Copenhagen, Jagtvej 128, DK-2200, Copenhagen, Denmark}

\author[0000-0001-7711-3677]{Claudia Maraston}
\affiliation{Institute of Cosmology and Gravitation, University of Portsmouth, Dennis Sciama Building, Burnaby Road, Portsmouth, PO1 3FX, UK}

\author[0000-0001-9189-7818]{Crystal L. Martin}
\affil{Department of Physics, University of California, Santa Barbara, Santa Barbara, CA 93109, USA}

\author[0000-0002-9489-7765]{Henry Joy McCracken}
\affiliation{Institut d’Astrophysique de Paris, UMR 7095, CNRS, and Sorbonne Universit\'{e}, 98 bis boulevard Arago, F-75014 Paris, France}

\author[0000-0002-6149-8178]{Jed McKinney}
\affiliation{The University of Texas at Austin, 2515 Speedway Blvd Stop C1400, Austin, TX 78712, USA}

\author[0000-0001-5846-4404]{Bahram Mobasher}
\affiliation{Department of Physics and Astronomy, University of California Riverside, 900 University Ave, Riverside, CA 92521, USA}

\author[0000-0002-4485-8549]{Jason Rhodes}
\affiliation{Jet Propulsion Laboratory, California Institute of Technology, 4800 Oak Grove Drive, Pasadena, CA 91001, USA}

\author[0000-0003-0427-8387]{R. Michael Rich}
\affiliation{Department of Physics and Astronomy, University of California Los Angeles, PAB 430 Portola Plaza, Los Angeles, CA 90095}

\author[0000-0002-1233-9998]{David B. Sanders}
\affiliation{Institute for Astronomy, University of Hawai’i at Manoa, 2680 Woodlawn Drive, Honolulu, HI 96822, USA}

\author[0000-0002-0000-6977]{John D. Silverman}
\affiliation{Kavli Institute for the Physics and Mathematics of the Universe (WPI), The University of Tokyo, Kashiwa, Chiba 277-8583, Japan}
\affiliation{Department of Astronomy, School of Science, The University of Tokyo, 7-3-1 Hongo, Bunkyo, Tokyo 113-0033, Japan}

\author[0000-0003-3631-7176]{Sune Toft}
\affiliation{Cosmic Dawn Center (DAWN), Denmark} 
\affiliation{Niels Bohr Institute, University of Copenhagen, Jagtvej 128, DK-2200, Copenhagen, Denmark}

\author[0000-0002-1905-4194]{Aswin P. Vijayan}
\affiliation{Cosmic Dawn Center (DAWN), Denmark} 
\affiliation{DTU-Space, National Space Institute, Technical University of Denmark, Elektrovej 327, 2800 Kgs, Lyngby, Denmark}

\author[0000-0003-1614-196X]{John R. Weaver}
\affil{Department of Astronomy, University of Massachusetts, Amherst, MA 01003, USA}

\author[0000-0003-3903-6935]{Stephen M.~Wilkins} 
\affiliation{Astronomy Centre, University of Sussex, Falmer, Brighton BN1 9QH, UK}
\affiliation{Institute of Space Sciences and Astronomy, University of Malta, Msida MSD 2080, Malta}

\author[0000-0002-8434-880X]{Lilan Yang}
\affiliation{Kavli Institute for the Physics and Mathematics of the Universe (WPI), The University of Tokyo, Kashiwa, Chiba 277-8583, Japan}

\author[0000-0002-7051-1100]{Jorge A. Zavala}
\affiliation{National Astronomical Observatory of Japan, 2-21-1 Osawa, Mitaka, Tokyo 181-8588, Japan}

\begin{abstract}
We report the discovery of 15 exceptionally luminous $10\simlt
z\simlt14$ candidate galaxies discovered in the first 0.28\,deg$^2$ of
{\it JWST}/NIRCam imaging from the COSMOS-Web Survey.  These sources
span rest-frame UV magnitudes of $-20.5>M_{\rm UV}>-22$, and thus
constitute the most intrinsically luminous $z\simgt10$ candidates
identified by {\it JWST} to-date.  Selected via NIRCam imaging with
{\it Hubble} ACS/F814W, deep ground-based observations corroborate
their detection and help significantly constrain their photometric
redshifts.  We analyze their spectral energy distributions using
multiple open-source codes and evaluate the probability of
low-redshift solutions; we conclude that 12/15 (80\%) are likely
genuine $z\simgt10$ sources and 3/15 (20\%) likely low-redshift
contaminants.
Three of our $z\sim12$ candidates push the limits of early stellar
mass assembly: they have estimated stellar masses
$\sim5\times10^{9}$\,\msun, implying an effective stellar baryon
fraction of
$\epsilon_{\star}\sim0.2-0.5$, where $\epsilon_{\star}\equiv
M_{\star}/(f_{b}M_{\rm halo})$.  The assembly of such stellar
reservoirs is made possible due to rapid, burst-driven star formation
on timescales $<$100\,Myr where the star-formation rate may far
outpace the growth of the underlying dark matter halos.  This is
supported by the similar volume densities inferred for
$M_\star\sim10^{10}$\,\msun\ galaxies relative to
$M_\star\sim10^{9}\,$\msun --- both about $10^{-6}$\,Mpc$^{-3}$ ---
implying they live in halos of comparable mass.  At such high
redshifts, the duty cycle for starbursts would be of order unity,
which could cause the observed change in the shape of the UVLF from a
double powerlaw to Schechter at $z\approx8$.  Spectroscopic redshift
confirmation and ensuing constraints of their masses will be critical
to understanding how, and if, such early massive galaxies push the
limits of galaxy formation in $\Lambda$CDM.
\end{abstract}

\keywords{}

\section{Introduction}\label{sec:intro}

The first year of {\it JWST} observations has revealed a wealth of
surprises, including the remarkable overabundance of luminous galaxies
in the epoch of reionization (EoR) relative to earlier expectation
\citep{bouwens15a,finkelstein16a,finkelstein22c,stark16a,robertson22a}.
      {\it Hubble}'s discoveries at $z>8$ told a story of a Universe
      rapidly growing at $z\sim9$ \citep{oesch18a}, yet very few
      candidates were identified at $z\sim10$. This suggested that
      galaxies grow in lock step with their halos with roughly equal
      star-forming efficiencies at all times (where `efficiency' is
      here defined as an effective stellar baryon fraction,
      $\epsilon_{\star} = M_\star / (f_{b}M_{\rm halo})$,
      where $M_\star$ is the stellar mass,
      $f_{b}=0.156$ the cosmic baryon fraction,
      \citealt{planck-collaboration20a}, 
        and $M_{\rm halo}$ the halo mass). Within
              the first few hundred million years (at $z>10$) the
              dearth of galaxy candidates from the pre-{\it JWST} era was
              considered to be a natural consequence of
              the halo-growth-limited conversion of baryons into stars
              \citep{bagley22a,harikane23a}, a process that was
              thought to be independent of redshift at early times
              \citep[e.g.][though some work has suggested its evolution,
                e.g. \citealt{coe13a,mcleod15a,mcleod16a,finkelstein16a,finkelstein22a}]{tacchella13a,mashian16a,stefanon17a,oesch18a,bouwens23a}.

Nevertheless, {\it Hubble}'s deepest surveys
led to the discovery of GN-z11 \citep{skelton14a,oesch14a,oesch16a},
then a $z\sim11$ candidate that was not only the most distant galaxy
candidate identified before JWST's launch, but also one of the most
luminous, with an observed rest-frame UV magnitude of
\muv$\,\sim-21.5$.  Selected from 0.2\,deg$^2$ of aggregate deep {\it
  Hubble} imaging, its implied volume density was
rather rare but difficult to constrain with a
single source.

We now know GN-z11 to be at $z=10.60$ thanks to {\it JWST} NIRSpec
observations \citep{bunker23a} with a star formation rate
$\sim$20\,\msun\,yr$^{-1}$ and stellar mass
$\sim$10$^{9}$\,\msun\ \citep{tacchella23a} --- all well in place
within the first 400\,Myr after the Big Bang.  Beyond its
extraordinary brightness, further {\it JWST} observations of GN-z11 reveal
even more surprises: it exhibits \lya\ in emission, has a
 possible damping wing of the
intergalactic medium observed as a Lorentzian absorption profile in
\lya\ \citep[and previously only seen as a signature of neutral IGM
  absorption in quasars,][]{miralda-escude98a}, as well as signatures
of a candidate accreting supermassive black hole, \lya\ halo, and
possible nearby companions \citep{scholtz23a}.  These observations
place new constraints on the early assembly of the highest density
peaks in the cosmic web.

Now that {\it JWST} is discovering new galaxies beyond $z>8$ by the
dozens \citep[if not
  hundreds;][]{finkelstein23a,harikane23a,franco23a}, we can directly
assess whether or not GN-z11 is so unique \citep{mason23a}.  Most new
discoveries have covered much fainter intrinsic luminosities with deep
NIRCam data obtained over somewhat narrow fields of view
($<$100\,arcmin$^2$), but the COSMOS-Web Survey
\citep[GO\#1727;][]{casey22a} is uniquely suited to the discovery of
bright, rare sources in the EoR.  In this paper we report the
discovery of several extremely bright candidate galaxies beyond
$z\simgt10$ found in the first 0.28\,deg$^2$ of NIRCam imaging data
from COSMOS-Web.  Though found in a similar survey area as the
0.2\,deg$^2$ of {\it Hubble} imaging used to find GN-z11, {\it Hubble}
could not have selected our candidates, as their detection relies on
the extraordinary depth provided by {\it JWST}'s long wavelength
imaging.  Throughout, we use a {\it Planck} cosmology
\citep{planck-collaboration20a}, AB magnitudes \citep{oke83a}, and a
Chabrier initial mass function \citep[IMF;][]{chabrier03a}.

\begin{figure}
  \includegraphics[width=0.99\columnwidth]{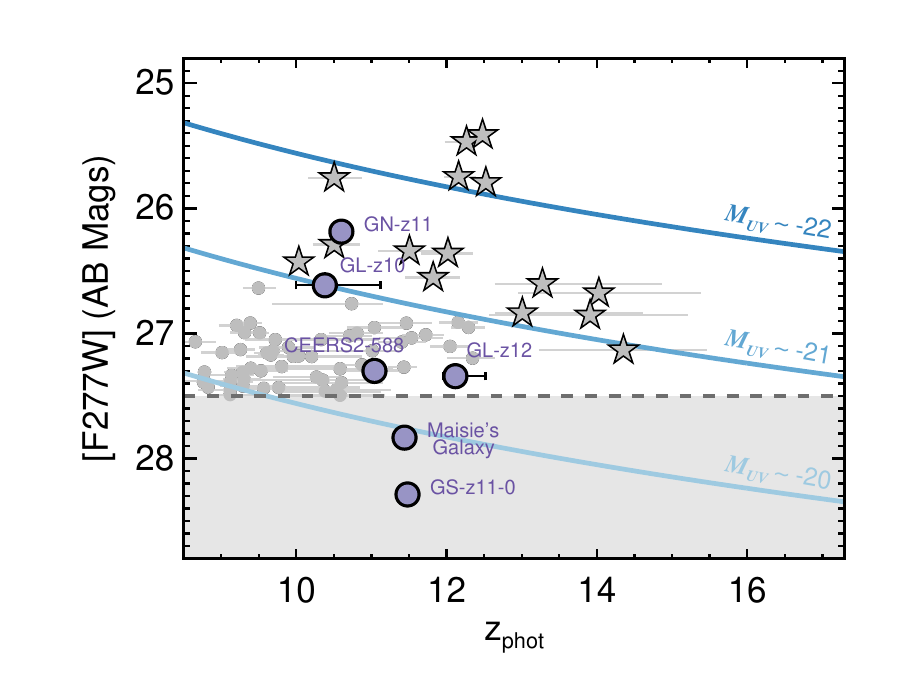}
  \caption{The distribution of bright candidate $z\simgt10$ galaxies
    we identify in the first 0.28\,deg$^2$ of COSMOS-Web (gray
    points).  The galaxies described in this paper (gray stars) are
    drawn from this sample, focusing on the particularly luminous
    subset (galaxies with initial estimates of \muv\,$<\,-21$). Note
    that the conversion from [F277W] to \muv\ as plotted is only
    approximate, as exact conversions depend on rest-frame UV
    slope. The gray region marks parameter space not explored in this
    work ([F277W]$>27.5$).  Well-known literature sources are shown in
    purple at their measured spectroscopic redshifts. The redshifts of
    GL-z10 and GL-z12 are given in accordance with their marginal
    detections of \oiii\ from ALMA \citep{bakx23a,yoon23a} and
    photometric uncertainties \citep{castellano22a}}.
\label{fig:magz}
\end{figure}

\section{Data}\label{sec:obs}

We select this sample of $z\simgt10$ candidates from the COSMOS-Web Survey
\citep[GO \#1727, PIs Kartaltepe \&\ Casey;][]{casey22a}, a 255\,hour
imaging program covering a contiguous 0.54\,deg$^2$ in four NIRCam
filters (F115W, F150W, F277W, and F444W); in parallel, observations
with MIRI in one filter (F770W) are obtained.  We refer the reader to
\citet{casey22a} for a detailed description of the survey design.  In
this paper, we explore the first two epochs of COSMOS-Web data taken
in January 2023 and April 2023, respectively.  In total, the area
surveyed in these two epochs is 0.28\,deg$^2$ for NIRCam and
0.07\,deg$^2$ for MIRI (covering 25\%\ of the NIRCam mosaic).

COSMOS-Web NIRCam data reduction was performed using the {\it JWST}
Calibration Pipeline \citep{bushouse22a} version 1.10.0, with the addition of several
custom modifications also implemented in other works
\citep[e.g.][]{bagley22a}.  This includes subtraction of 1/f noise and
background.  The Calibration Reference Data System (CRDS) pmap-1075
was used, corresponding to NIRCam instrument mapping imap-0252.
Astrometry is anchored to Gaia-EDR3 and bootstrapped from the {\it
  Hubble}/F814W imaging \citep{koekemoer07a} and COSMOS2020 catalogs
\citep{weaver22a}.  The normalized median absolute deviation on
astrometry is less than 12\,mas for all filters.  Mosaics with a
30\,mas pixel scale are produced in stage 3 of the pipeline.  A
forthcoming paper (M. Franco \etal) will provide a more complete
description of the COSMOS-Web NIRCam image processing.  Similarly, we
reduce MIRI data using the same pipeline with similar custom
modifications; a more complete description of COSMOS-Web MIRI imaging
(S. Harish \etal) will follow.

Beyond JWST, we use the wealth of multiwavelength data in the COSMOS
field to vet detections presented in this paper, from the {\it
  Hubble}/F814W imaging from the original COSMOS Survey
\citep{scoville07a,koekemoer07a}, the {\it Spitzer} COSMOS Survey
\citep{sanders07a}, Subaru Telescope Hyper Suprime-Cam (HSC) imaging
\citep{aihara22a}, and UltraVISTA imaging \citep{mccracken12a},
updated to the most recent release DR5.  A full description of these
datasets is provided in \citet{weaver22a} and
\citet{casey22a}\footnote{COSMOS2020 used UltraVISTA DR4 imaging,
while this work uses
\href{https://www.eso.org/sci/publications/announcements/sciann17572.html}{UltraVISTA
  DR5}.}.  We note that none of the targets included in our analysis
are detected in the COSMOS2020 photometric catalog \citep{weaver22a},
which used a deep $\chi^2$ detection image constructed using
UltraVISTA $YJHKs$ plus HSC $iz$ bands to extract source photometry;
the $z>7.5$ sources identified in COSMOS2020 were limited to those
presented in \citet{kauffmann22a}, which have similar rest-frame UV
luminosities ($M_{\rm UV}<-21$) to those presented herein, but
are generally at lower redshifts $7<z<10$ and identified over a wider
area.
Similarly, none of our sources have significant emission in the
millimeter \citep[4 of 12 sources have some ALMA coverage, while all
  are covered by deep single-dish data from SCUBA-2;][McKinney \etal,
  in prep]{simpson19a}, X-ray \citep{civano16a}, or radio
\citep{jarvis16a,smolcic17a}.

 We note that MIRI data covers only 0.07\,deg$^2$ of the 0.28\,deg$^2$
 covered to-date in COSMOS-Web (25\%); of the 15 candidates we discuss
 in this paper (whose selection is outlined in the next section), only
 3 are covered by MIRI 7.7\um\ pointings.  Of those, only one is
 detected: \galc\ at 7.8$\sigma$ (the measured flux density given
 later in \S~\ref{sec:details}).  We note that detection at this
 threshold, as well as non-detection, is not particularly constraining
 to the SED fits though the constraints are included in our
 analysis. Had any candidates been detected at 7.7\um\ at higher
 significance ($\simgt100\sigma$), it would suggest they are more
 likely to be lower-redshift contaminants.

\section{Photometry, Source Selection, \&\ Measurements}\label{sec:selection}

Photometric catalogs for COSMOS-Web imaging were constructed using the
model-based photometric package {\tt SourceXtractor++} ({\tt SE++};
\citealt{bertin20a,kummel20a}).  A detection image is constructed
using a $\chi^2$ combination
\citep{szalay99a} of all four NIRCam filters (F115W, F150W, F277W, and
F444W), which is then used for reference to extract photometry from
each individual band.  Extraction is carried out on {\it JWST}
imaging, {\it Hubble} imaging, and all ground-based datasets described in
\citet{weaver22a}. Model-based photometry allows simultaneous
photometric measurements to be made on images with different point
spread functions (PSFs) without degradation.  This allows us to fold
in constraints from the deep ground-based data without loss of
resolution (thus photometric precision) in our space-based data.

Alongside the primary {\tt SE++} photometry, we use {\tt Source
  Extractor} ({\tt SE}) `classic' \citep{bertin96a} to perform
aperture photometry on PSF-homogenized images from {\it Hubble} and
{\it JWST}.  PSF homogenization is performed using
empirically-measured PSFs built with PSFEx \citep{bertin11a} and the
{\tt pypher} package \citep{boucaud16a} that computes a homogenization
kernel between two different PSFs.  All images are PSF homogenized to
the F444W image.  The same detection image is used for {\tt SE}
`classic' as was for {\tt SE++}.  A number of tests were performed to
check the consistency of model-based and aperture-based photometry,
all of which will be described in a forthcoming paper (M. Shuntov
\&\ L. Paquereau \etal).  Allowing full flexibility in the {\tt SE++}
catalog construction leads to some underestimated errors for faint or
undetected sources in certain bands; we address this by setting a
floor to the noise estimates in each filter equal to the shot noise
measured in random circular apertures of diameter 0$\farcs$3 (for {\it
  Hubble} and {\it JWST} imaging) and 1$''$ (for ground-based
imaging).  For the purposes of this work, we adopt model-based
photometry but also provide aperture-based measurements for reference
to the reader.

All sources in the {\tt SE++} and `classic' catalogs are then fit to
photometric redshifts using the {\tt EAzY} SED-fitting tool
\citep{brammer08a} with a combination of the default Flexible Stellar
Population Synthesis (FSPS) templates \citep[][specifically QSF 12
  v3]{conroy10b} and bluer templates optimized for selecting less
dusty galaxies in the EoR from \citet{larson22b}, which
  makes a number of models available with variable Ly$\alpha$ escape
  fractions.  For our initial runs we adopt the reduced Ly$\alpha$
template set; we run {\tt EAzY} a second time on candidate $z>10$
galaxies using the template set without Ly$\alpha$, and perceive no
change in the resulting redshift probability density distributions (PDFs).
We do not use a magnitude prior in our {\tt EAzY} runs, thus adopt the
default flat redshift prior.

The initial identification of candidate $z>10$ galaxies is performed
using the following criteria on the {\tt SE++} model-based photometric
catalog:
\begin{enumerate}
  \vspace{-1mm}
\item ${\rm SNR_{F814W}<2}$, ${\rm SNR_{F115W}<2}$, ${\rm
  SNR_{F277W} \ge 5}$, and ${\rm SNR_{F444W} \ge 5}$,
  \vspace{-1mm}
\item Best-fit photometric redshift from {\tt EAzY} of $z_{\rm a}\ge8.5$,
  \vspace{-1mm}
\item The integral of the redshift probability distribution above
  redshift $6$ is $p(z>6)\ge0.95$ and above redshift $8$ is
  $p(z>8)\ge0.80$, and
  \vspace{-1mm}
\item Magnitude in F277W is ${\rm [F277W] \le 27.5}$.
  \vspace{-1mm}
\end{enumerate}
Of $\sim$340,000 sources identified in the 0.28\,deg$^2$, 620 sources
fulfill this criteria, of which 340 (55\%) are identified as hot
pixels and are thrown out using size threshold criteria
(i.e. sources with {\tt SE++} area less than 100 modeled
  pixels and radius smaller than 0.01\,arcsec).  All remaining 280
candidates are visually vetted.  Because we did not directly select
sources using {\it JWST} color cuts, many are somewhat red
([F277W]--[F444W]\,$>1$) as well as spatially extended.
  As a result they are likely to be at lower
redshifts than $z\sim10$ and thus removed; 86 sources remain as
plausible $z\simgt10$ candidates.  Their distribution in observed
[F277W] magnitude against photometric redshift is shown in
Figure~\ref{fig:magz} including some
other well-known high-$z$ sources in the literature including GN-z11
\citep{bunker23a,tacchella23a}.
From these 86, we closely scrutinize sources that are 
 estimated to have absolute UV magnitudes brighter than
$M_{\rm UV}\sim-21$ and $z\ge10$ as shown in
Figure~\ref{fig:magz}. There are fifteen clearly
  separated from the population of fainter sources found at similar
  redshifts in COSMOS-Web.  Note that all 15 sources
  also pass the SNR criteria of our selection using {\tt SE}
  `classic' aperture photometry.  We limit analysis to this subsample
only in this paper, focusing on galaxies at the most extreme
margins of luminosity and redshift.

We then apply more stringent fits to the fifteen $z\simgt10$ candidates in
our dataset.  First, we repeat the photometric redshift fitting of all
candidates using {\tt LePhare} \citep{arnouts02a,ilbert06a}, {\tt
  Bagpipes} \citep{carnall18a},
and {\tt EAzY}.  We generate optimal fits after running each tool
twice: once with a flat redshift prior from $0<z<20$ and second with a
flat prior from $0<z<7$ to generate best-fit low redshift template
fits.  All codes use full flux density constraints and
  their uncertainties in all broadband filters (rather than upper limits).

Our {\tt EAzY} runs are slight modifications of the initial tests used
to select the sample: we allow a wider range of possible redshift
solutions with finer redshift sampling, and shift the adopted template
set from \citet{larson22b} to those with no Ly$\alpha$ emission to
account for a presumably neutral IGM at $z\simgt10$.

For {\tt LePhare} optimal fits, we follow the methodology of
\citet{kauffmann22a} and briefly summarize here.  \citet{bruzual03a}
templates are used spanning a range of star formation histories
(exponentially declining and delayed) as in \citet{ilbert15a}, with
two different dust attenuation curves \citep{calzetti00a,arnouts13a}.
Emission line fluxes are added following \citet{saito20a}, allowing
variation in line strength by a factor of 0.3\,dex from expectation
\citep{schaerer09a}.  Note that Ly$\alpha$ emission is included in
these fits, though not at significantly high equivalent widths to
impact the fits to the photometry of these bright sources.

For {\tt Bagpipes} optimal fits, we implement a
delayed-$\tau$ star formation history as a fraction of the age of the
universe\footnote{This is done to ensure that a flat prior is assumed
as a function of redshift rather than a flat prior on the age of the
stellar population, which is automatically capped at the age of the
Universe at any given redshift.} at redshift $z$ plus a
  recent, instantaneous burst lasting between 1--100\,Myr. We use a
Calzetti dust attenuation law \citep{calzetti00a} with stellar
population models from \citet{bruzual03a}.  We allow the absolute
magnitudes of attenuation $A_{V}$ to span 0--3  to
  capture a reasonable range of attenuation, ranging from unobscured
  to values more heavily attenuated than seen in typical submillimeter
  galaxies' integrated light \citep[e.g.][]{da-cunha15a}.  We note
that restricting to $A_{V}<3$ is required to prevent fits using rather
unphysical models, for example extremely dust-attenuated dwarf
galaxies with $A_{V}\sim6$ at low redshift ($z<1$). We later
  discuss why such low-mass, extreme attenuation models are
  inconsistent with the sample.

We also fit photometry to model libraries of brown dwarf SEDs from
\citet{morley12a} and \citet{morley14a}, spanning temperatures
200\,K--2000\,K and a range of surface gravities; none of our sample
are well fit by brown dwarf templates.  As discussed in
  \S~\ref{sec:other}, all sources in our sample are also spatially
  resolved, further reinforcing the idea that it is unlikely any are
  brown dwarfs.

We then repeat a number of these fits on the photometry extracted from
the {\tt SE} `classic' 0$\farcs$3 diameter apertures on all
constraining space-based imaging (i.e. {\it Hubble}/F814W and the {\it
  JWST}/NIRCam filters). We determine that differences in measured
photometry largely do not impact the results, though some nuances of
the differences (and their effects on the redshift probability
distributions) are discussed again later in \S~\ref{sec:lowz}.  We
adopt the model-based photometry from {\tt SE++} as our fiducial
photometry and find that all fits (with {\tt EAzY}, {\tt
    LePhare}, and {\tt Bagpipes}) estimate $<$3\%\ of the redshift
  probability distributions is at $z<7$. 
  
In this work, we adopt the posterior distributions of physical
parameters from our {\tt Bagpipes} tests to describe the
characteristics of the sample, such as stellar mass, 100\,Myr-averaged
star formation rates, \muv, the rest-frame UV slope $\beta$, and the
absolute magnitude of attenuation $A_{\rm V}$.  The motivation for
such a choice is further discussed in \S~\ref{sec:other}.  In
addition, {\tt Bagpipes} is the only code that directly provides
posterior distributions in all physical parameters, giving valuable
insight on covariances and the nature of potential contaminants.  For
sources with significant fractions of their redshift solutions fit beyond
$z>15$ (those presented in \S~\ref{sec:z13}), we enforce an artificial
cap at $z\sim15$ for their physical characteristics, realizing
solutions between $15<z<20$ are far less likely than $13<z<15$.  We
show all of the SED fits for the sample of 15 candidates in
Figure~\ref{fig:bio1}.

\subsection{Quantifying Goodness of Fit}\label{sec:chi2}

For each source, we quantify a normalized $\chi^2$ metric,
which we calculate as $\chi^2_{n}=\chi^2/N_{\rm bands}$, where $N_{\rm
  bands}$ is the number of effective bands available to us that are
most directly constraining for $10<z<14$ galaxies. Here we adopt
$N_{\rm bands}=7$, taking the five space-based bands ({\it Hubble}
F814W, and {\it JWST} F115W, F150W, F277W, and F444W) and we count two
ground-based filters whose depths, $\sim$27.5-28.2 AB at 5$\sigma$
\citep{casey22a}, are most useful for this work: UltraVISTA $H$ and
$Ks$.  A future work on the ultraviolet luminosity function (UVLF)
from COSMOS-Web will explore a more careful, quantitative definition
of $\chi^2_{n}$.  \input{biofigure} \input{tab_phot}

\noindent Note that $\chi^2_{n}$ is {\it not} a reduced $\chi^2$, which would
also account for the degrees of freedom in model fits\footnote{A
subtlety of $\chi^2_{n}$ is that it may dip below unity. While the
same behavior for a reduced $\chi^2$ would be an indication of an
overfit dataset, that is not the case for $\chi^2_{n}$ where we are
not accounting for the number of free parameters in the model (if we
were, the value of $\chi^2_{n}$ would increase).} That is 
a complex
problem in SED fitting, as often models (used
to
  generate templates in SED fitting) have more tunable
parameters than galaxies have photometric detections, 
and reducing the
searchable parameter space by making well-motivated physical choices
differs greatly between SED fitting tools.

We motivate such a normalized $\chi^2$ to have a quantity that could
be directly comparable between surveys that have a variety of
different filters used to select their samples. For example, galaxies
selected in the CEERS survey \citep{finkelstein23a} are constrained
with more deep space-based photometric bands, and would naturally have
a higher value of $\chi^2$, thus larger differential values
$\Delta\chi^2$ between low and high redshift solutions.  Normalizing
by the number of constraining bands would make selection of sources in
these different surveys directly comparable.  For example, the
oft-used $\Delta\chi^2>4$ criterion
\citep[e.g.][]{finkelstein23a,hainline23a} here would translate to
$\Delta\chi^2_{n}\simgt0.6$ given the 7 bands that are constraining
for COSMOS-Web $z>10$ candidates.  Given the unique nuances of each
redshift fitting tool, we only directly compare $\chi^2_{n}$ for
low-$z$ and high-$z$ fits using the same fitting code (e.g. {\tt EAzY}
low-$z$ to {\tt EAzY} high-$z$, {\tt LePhare} low-$z$ to {\tt LePhare}
high-$z$ and {\tt Bagpipes} low-$z$ to {\tt Bagpipes} high-$z$).  Thus
the criteria for a robust candidate is if
$\Delta\chi^2_{n}\equiv\chi^2_{{\rm n,low-}{\it z}}-\chi^2_{{\rm
    n,high-}{\it z}}>0.6$.

\input{tab_phys}

For transparency in our selection, Figure~\ref{fig:bio1} shows {\it
  all} 15 $z>10$ candidates with $M_{\rm UV}\simlt-21$, though some of
them have viable low-redshift solutions as measured via
$\Delta\chi^2_{n}$.
This includes \galz, which we have multiple reasons to believe is at
low-$z$ (detailed in \S~\ref{sec:details}), as well as \galk\ and
\gali, which are only detected in the F277W and F444W bands, discussed
further in \S~\ref{sec:z13}.  Note that sources detected in only two
bands are naturally more difficult to cleanly select using a $\chi^2$
metric as most data constraints are upper limits.  Nevertheless, these
sources are discussed individually in \S~\ref{sec:z13} and again in
\S~\ref{sec:lowz} regarding the probability that they are low-redshift
interlopers.

\subsection{Other measurements}\label{sec:other}

We conducted a simple test to assess the accuracy of the photometric
redshift estimates using the  suite of  best-fit SEDs
for the whole sample from {\tt EAzY}, {\tt LePhare} and
{\tt Bagpipes}. This set of templates is assumed to
  sample the breadth of realistic SEDs for $z>10$ galaxies.  We
shifted these SEDs forward and backward in redshift, modeled synthetic
photometry with characteristic noise of our data, and remeasured
photometric redshifts.  We found a systematic offset towards higher
redshifts using {\tt EAzY}, while {\tt Bagpipes} exhibited no
systematic offset.  This is similar to the {\tt
    Bagpipes} and {\tt EAzY} systemic offsets seen in the first-epoch
  COSMOS-Web $z\sim9-11$ galaxies from \citet{franco23a}.  We adopt
the physical parameters and redshifts measured by {\tt Bagpipes} for
the rest of this work.

We measure sizes of the sample using the F277W band using both {\sc
  Galfit} \citep{peng02a,peng10a} and {\sc Galight} \citep{ding20a}.
  The F277W band is chosen as it provides the highest S/N
measurements for all galaxies in the sample with the best spatial
  resolution. {\sc Galfit} uses a least-squares fitting algorithm
    while {\sc
    Galight} uses a forward modeling approach to find the optimum
S\'ersic fit to a galaxy's two dimensional light profile, after
accounting for the PSF.  We use average PSF images generated from our
April 2023 epoch of data measured using PSFEx \citep{bertin11a}.  The
size measurements were broadly consistent between the two fitting
techniques and in all cases spatially resolved; we provide the
half-light radius measurements from {\sc Galfit} in
Table~\ref{tab:phys}.

\begin{figure*}
  \centering
  \includegraphics[width=0.95\textwidth]{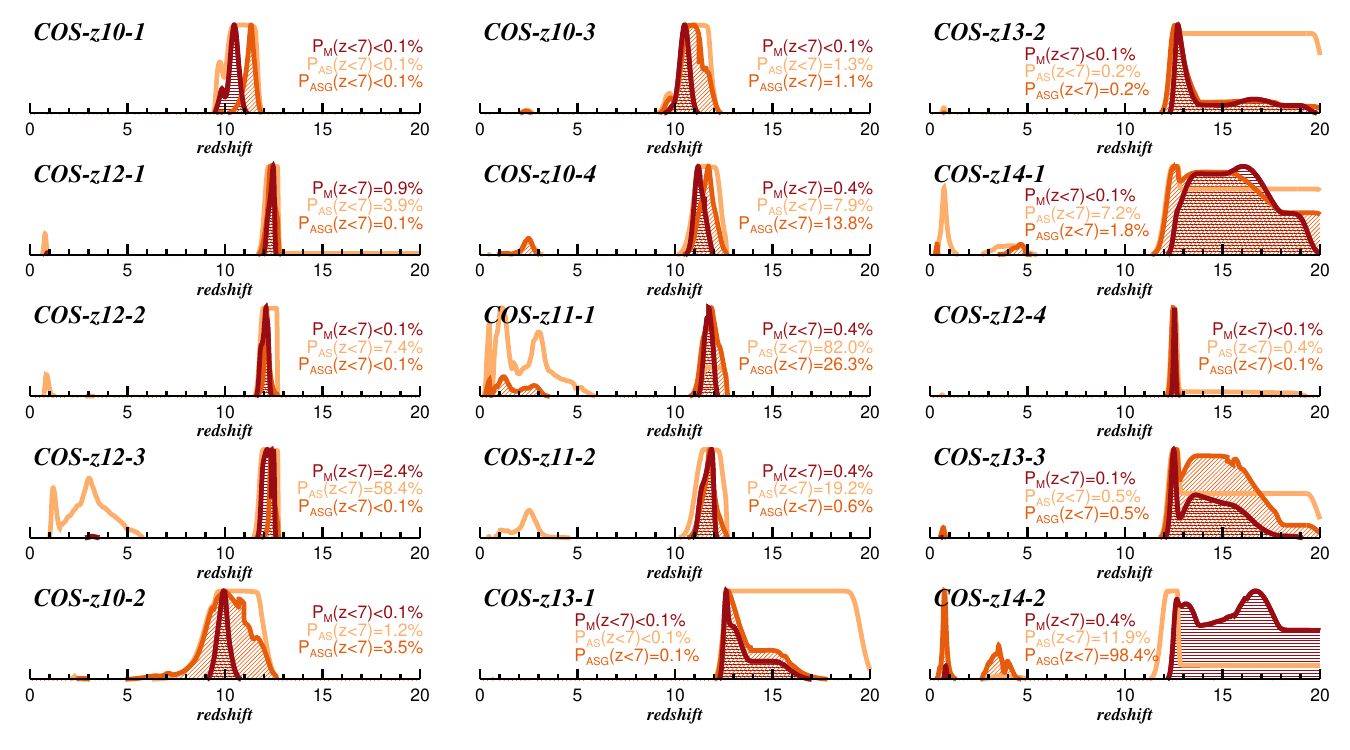}
  \caption{A comparison of redshift probability density distributions
    fit using {\tt EAzY} to three sets of photometry for each high-$z$
    galaxy candidate.  The results for the {\tt SE++} model-based
    photometry (`M'), which includes both space and ground-based
    photometry, are shown in dark red.  Light orange shows the results
    for the {\tt SE} classic photometry extracted in 0$\farcs$3
    diameter apertures from PSF-homogenized {\it Hubble} and {\it JWST}
    NIRCam images only (`AS' for Aperture, Space-based only). Dark
    orange shows the distributions after joining together the
    aperture-based photometry from {\tt SE} classic with the
    model-based photometric measurements of ground-based data (`ASG',
    Aperture with Space and Ground).  This last fit is meant to
    highlight the relative value and importance of the ground-based
    constraints, particularly those from deep Subaru/HSC imaging in
    the optical and UltraVISTA imaging in the near-infrared.  Inset
    are the percentages of the redshift PDFs that lie at $z<7$.  In
    the majority of cases, the model-based photometry and
    aperture-based photometry produce similar redshift PDFs and the
    addition of the ground-based constraints to the aperture-based
    photometry dramatically reduce the integral of the redshift PDF
    below $z<7$ for most sources.}
  \label{fig:zpdfs}
\end{figure*}

\section{Source Details}\label{sec:details}

Here we provide more detailed descriptions and relevant information
about each of the 15 candidate galaxies at $z>10$.  We note that the
sample may be delineated in roughly three subsets: exceptionally
luminous $10<z<12$ candidates (\muv\,$<$\,$-21.5$), bright $10<z<12$
candidates ($-20.5>$\,\muv\,$>-21.5$), and $z>13$ candidates with
\muv\,$<-20.5$.  These three samples contain five sources each.  The
$z>13$ candidates are only detected in two {\it JWST} bands: F277W and
F444W and thus are not as well constrained in terms of their physical
properties.  Photometry is provided for all sources in
Table~\ref{tab:phot} and positions and derived physical properties are
given in Table~\ref{tab:phys}.

In the descriptions that follow, we quote the relative probability
that a source is a low-redshift interloper; we draw these
probabilities from the {\tt EAzY} photometric redshift fits using the
{\tt SE++} photometry, whose distributions are presented in
Figure~\ref{fig:zpdfs} and later discussed as an ensemble in
\S~\ref{sec:lowz}.  We chose {\tt EAzY} redshift PDFs due to their
simplicity and straightforward adoption of a flat redshift (and
magnitude) prior.  However, a general caveat of interpreting redshift
PDFs is that the exact amplitude of redshift peaks are quite sensitive
to the adopted template set, range of physical parameters governing
the input SEDs, and to small differences in adopted photometry.  So
while generally multiple independent codes find consist peaks in the
distribution (also shown for inset panels in Figure~\ref{fig:bio1}),
the integral under each peak is somewhat uncertain.

We also address the presence of on-sky neighbors as possible physical
associations, checking for consistency with low-redshift solutions to
our candidates \citep[particularly because close neighbors on the sky
  are statistically more likely to be physically
  associated;][]{kartaltepe07a,shah20a}.

\subsection{Exceptionally Luminous $10\simlt z\simlt 12$ Galaxies}\label{sec:superbright}

\subsubsection{\gala}

This source is detected in five bands and exhibits a 2.5 magnitude
drop between F150W and the 2$\sigma$ upper limit in F115W.  \gala\ is
formally detected in UltraVISTA $H$ and $Ks$ band imaging in addition
to the three reddest NIRCam filters. Its redshift is fit to $z_{\rm
  phot}=10.3-10.4$ using {\tt EAzY} and {\tt LePhare} and slightly
lower at $z_{\rm phot}=9.7$ from {\tt Bagpipes}.  \gala\ has two
neighbors within 1$\farcs$5: one 0$\farcs$47 to the southwest and
another 1$\farcs$21 to the south-southwest.  The former has a
photometric redshift of $z_{\rm phot}=0.97$, and the latter $z_{\rm
  phot}=2.04$, both of which are inconsistent with the forced
low-redshift solutions for \gala\ at $z_{\rm low}=2.4$, indicating
that physical association with the neighbors (adopting the
low-redshift solution) is unlikely. We estimate the maximum lensing
that could occur from these foreground sources as $\mu\approx$1.01
given the estimated masses of the foreground objects
$\simlt$10$^{8}$\,\msun. There is no significant low-redshift peak in
the redshift probability density distribution at low redshift for
\gala, whose integrated probability of being at $z<7$ is $<$0.1\%.

\subsubsection{\galb}

This galaxy is detected at high signal-to-noise in three bands (F277W,
F444W and UltraVISTA $Ks$) with a substantial 5$\times$ flux density
drop (1.7 magnitudes) in the F150W filter, and another factor of 4
drop (1.5 magnitudes) to the 2$\sigma$ upper limit in F115W.  \galb\ does not have an abrupt
spectral break, though its photometry can be explained well with a
redshift $z=12.0-12.5$.  Forced low-redshift solutions produce
photometric redshift estimates at $z_{\rm low}<1$ for {\tt EAzY} and
{\tt LePhare} and at $z_{\rm low}=3.2$ for {\tt Bagpipes}.  Solutions
at $z_{\rm low}<1$ would require extreme emission line strengths for a
relatively low-mass galaxy
  (sSFR$\simgt10^{-7}$\,yr$^{-1}$), thus are less likely
than the $z_{\rm low}=3.2$ {\tt Bagpipes} solution, though all are
significantly less well fit to the data than the high-redshift
solutions ($p(z<7)\sim0.9$\,\%).  A foreground neighbor, located
1$\farcs$2 to the southeast, is fit to a photometric redshift of
$z_{\rm phot}=2.55$ but sufficiently distant to not contaminate
\galb's photometry; its photometry and $z_{\rm phot}$ is
  also distinct from \galb's low-redshift solutions to not suspect
  physical association.  \galb\ is the most intrinsically luminous
$z>12$ galaxy found in our sample with
\muv\,$=-22.19^{+0.10}_{-0.17}$.

\subsubsection{\galc}

Detected in five bands (F150W, UltraVISTA $H$ and $Ks$, F277W and
F444W), \galc\ has a factor 4 drop in flux density (1.5 magnitudes)
between UltraVISTA $H$ and F150W which demarcates the candidate's
Lyman break.  Its photometric redshift estimates span $z_{\rm
  phot}=11.9-12.1$ with possible low-redshift solutions $z_{\rm
  low}=1-3$.  The low-redshift solution that is most plausible is
likely the strong-line emitter at $z=3.03$ found with {\tt LePhare},
though not quite as extreme a drop around $\lambda=1.5$\,\um\ would be
expected compared to observations.  There are no close neighbors
within 1$\farcs$5 of \galc, rendering its photometry clean from
contamination.  There is a source 1$\farcs$9 distant to the northwest
that has a photometric redshift from COSMOS2020 of $z_{\rm
  phot}=3.68$, which is closer to but still statistically distinct
from \galc's low-redshift solution.  We note that, depending on the
exact tuning of template sets for {\tt EAzY} or the adopted range of
physical parameters used in {\tt Bagpipes}, as much as 2\%\ of the
redshift probability density distribution sits at $z\sim3$ (and it
goes as high as 7.4\%\ using only aperture photometry on space-based
constraints).  This makes \galc\ slightly less robust as a $z>10$
candidate than others in this exceptionally luminous category.  We
note that \galc\ is also detected with MIRI at 7.7\um, the only galaxy
in our sample to have such a detection; its flux density is $S_{\rm
  7.7\mu m}=288\pm37\,$nJy; this detection is consistent with an
approximately flat spectrum (in $F_{\nu}$) from the near-IR.
\galc\ is the second brightest galaxy in this sample.

\subsubsection{\gald}

This source is detected in five bands (F150W, UltraVISTA $H$ and $Ks$,
F277W and F444W) with a strong factor of 5 drop from UltraVISTA $H$ to
F150W.  \gald\ has a redder rest-frame UV slope than other galaxies in
this sample, which introduces more possible degeneracies with a
low-redshift origin to its photometry.  Nevertheless, the strong break
at $\lambda\sim 1.5$\,\um---a factor of 6--10 in flux density (2--2.5
magnitudes)---argues for a high-redshift solution, and its
photometric redshift is consistently fit to $z_{\rm phot}=11.5-12.2$.
Without the UltraVISTA photometry, \gald\ would be difficult to
identify.  \gald\ has no neighbors within 3$\farcs$0.

The inferred absolute UV magnitude for \gald\ is $M_{\rm
  UV}=-21.58^{+0.07}_{-0.31}$, and its dust-corrected SFR of
$54\pm16$\,\sfr\ is the highest of the sample.  With a rest-frame UV
slope of $\beta=-0.60^{+0.22}_{-0.33}$ and
$A_{V}=0.58^{+0.22}_{-0.33}$, it is the reddest of the bright sources
in this paper; this combined with the high SFR leads naturally to a
hypothesis that it may be detected (or detectable) in the millimeter.
\gald\ is one of the few sources covered by existing archival ALMA
data at 2mm from program (2021.1.00705.S, PI: O. Cooper); it is
undetected with measured RMS of $\sim$0.08\,mJy, which sets an
approximate 2$\sigma$ upper limit to the dust mass at $z=11.5$ of
4$\times$10$^8$\,\msun\ and $L_{\rm
  IR}\simlt6\times10^{11}$\,\lsun\ (SFR\,$\simlt$\,90\,\msun\,yr$^{-1}$);
both limits are not sufficiently constraining to be in conflict with
the measured $A_{V}$ from the {\tt Bagpipes} fit.  Detecting
millimeter continuum in such sources may require 2\,mm observations
with a sensitivity $\sim$0.01\,mJy, though we stress that due to the
negative K-correction, dust continuum observations of such sources
does not confirm or refute a low- or high-redshift solution;
spectroscopy is necessary.

\subsubsection{\galz}

The fifth exceptionally-bright candidate we identify is \galz.  The
primary limitation in characterizing \galz\ is the proximity of two
neighbor galaxies whose emission is spatially confused in ground-based
data.  While nominally fit to a photometric redshift $z_{\rm
  phot}=12.3-12.7$ and $M_{\rm UV}=-21.90^{+0.15}_{-0.15}$, the {\tt
  SE++} model-based measurements for \galz\ claim a 3$\sigma$
detection in the HSC-$z$ band, and imaging from SuprimeCam $i$-band
may show a $\sim$1.5$\sigma$ detection; however, on close inspection,
both may be positive noise fluctuations from the neighbors'
emission. The lack of deep, optical constraints with high-resolution
imaging deems the source less secure.  The neighbors are 0$\farcs$47
away to the west and 1$\farcs$0 away to the southwest and are fit to
photometric redshifts of $z_{\rm phot}=4.65$ and $4.50$
respectively. The former is also detected in COSMOS2020 with a
consistent photometric redshift of $z_{\rm phot}=4.8$.  Indeed, the
{\tt LePhare} low-redshift photometric fit to \galz\ is consistent
with the neighbors' redshifts, at $z_{\rm low}=4.65$.  Though the best
high-$z$ fit still has a formally lower $\chi^2_{n}$ than this low-$z$
solution, the consistent photometry with a low-$z$ neighbor is
sufficient to cast significant doubt on the high-$z$ solution. The
situation is similar to the environmental coincidence that argued
CEERS-93316 was $z\sim4.9$ and not $z\sim16$ from \citet{naidu22b},
later confirmed by \citet{arrabal-haro23b}.  We therefore choose to
remove \galz\ from our analysis for the rest of this paper.
Nevertheless, we provide its measured physical characteristics {\it
  if} it were at $z>10$ in Table~\ref{tab:phys}.  We emphasize that
obtaining a spectrum of \galz\ is important in case it is indeed an
ultra high-redshift source: in that case we will have learned a
valuable lesson about the claimed completeness of our survey, the true
occurrence rates of chance projections with low-$z$ sources, and the
abundance of ultra-bright $z\sim12$ sources.

\subsection{Bright $9<z<13$ Candidates}\label{sec:bright}

\subsubsection{\galn}

\galn\ is one of the more intrinsically red galaxies in our sample and
has a slightly bimodal morphology in the NIRCam LW bands.  It is
formally detected in three NIRCam bands with marginal detections in
UltraVISTA $J$, $H$, and $Ks$.  The marginal $J$-band detection
combined with the deep upper limit from F115W pegs the redshift right
around $z\sim9$ at the low redshift end of our sample.  The
photometric redshift estimates fall between $z=9.1-10.1$.  The derived
rest-frame UV slope for \galn\ is $\beta=-$1.22$^{+0.22}_{-0.16}$, the
second reddest of the sample; despite its red color, it passes all
$\Delta\chi^2_{n}$ criteria thanks to the strength of its Lyman-break,
which is a factor of 6 in flux (2 magnitudes).  The source has no
neighbors, but its double-component morphology may cast some doubt on
the reliability of single S\'ersic profile-based measured photometry.
However, the aperture-photometry for this source is fully consistent
with the model-based results (albeit with more uncertainty in the
redshift PDF, see Figure~\ref{fig:zpdfs}).  \galn\ has the highest
derived stellar mass estimate of any in our sample with
$M_\star=(1.1^{+0.7}_{-0.3})\times10^{10}$\,\msun\ (a consequence of
its redder color in the rest-frame UV).

\subsubsection{\gale}

\gale\ is detected in three NIRCam bands; though formally undetected
($<$3$\sigma$) in UltraVISTA, cutouts clearly show excess emission
consistent with the 2$\sigma$ upper limit in $H$-band.  \gale\ has a
close neighbor 0$\farcs$78 to the southeast that has a photometric
redshift of $z_{\rm phot}=2.32$ that is also found in COSMOS2020 with
a similar photometric redshift.  We note that this is somewhat
consistent with one of the three low-redshift solutions for
\gale\ found by {\tt EAzY} of $z=2.37$, which raises the probability
this is a low-redshift interloper.  However, we note that such a
solution is a relatively poor fit compared to its corresponding
high-redshift solution ($\Delta\chi^2_{n}=0.9$), sufficiently high to
remain in the sample.

\subsubsection{\galf}

\galf\ is similarly detected in three NIRCam bands with a compact core
and diffuse extended emission. There is no indication that it is
detected in UltraVISTA imaging. The stack of HSC $griz$ imaging
displays a puzzling excess emission to the south-west; because this is
not spatially coincident with the source's position within 1$''$, it is not of significant
concern. \galf\ has no close neighbors, and has redshift solutions
spanning $z_{\rm phot}=10.2-11.2$, with significantly better high-$z$
fits than forced low-$z$ solutions.

\subsubsection{\galh}

\galh\ is the highest redshift of the `bright' subset and is detected
in the three NIRCam bands only, with detections in F150W, F277W and
F444W.  There is no evidence from the ground-based cutouts of
significant emission above the background noise. It has no neighbors
and is fit to redshifts $z_{\rm phot}=11.2-11.7$.  Using model-derived
photometry, 0.4\%\ of the redshift PDF is at $z<7$, though a
significantly higher $z<7$ probability is found while using
aperture-based photometry alone (62\%).  However, adding the
ground-based constraints in with the aperture-based photometry, the
$z<7$ solutions occupy a lower percentage of the distribution at
26.3\%.

\subsubsection{\galg}

\galg\ is detected in three NIRCam bands plus UltraVISTA $H$-band,
such that a clear break is detected between UltraVISTA $H$ and
F150W. It lacks neighbors and has redshift solutions spanning
$z=10.6-11.9$.  The low-redshift solutions for this source cluster
around $z=2.1-2.6$ and are only marginally worse
($\Delta\chi^2_{n}=0.3-0.5$) than the high redshift solutions, likely
because it is both relatively faint and redder than other sources in
the sample, with $M_{\rm UV}=-20.77^{+0.23}_{-0.34}$ and
$\beta=-1.48^{+0.27}_{-0.39}$.  The low-redshift solutions occupy
$\sim$0.4\%\ of the redshift probability distribution, though we note
that adopting the aperture-based photometry results in a larger
fraction of the redshift PDF at $z<7$: 7.9\%\ using aperture-based
photometry alone, and 13.8\%\ using aperture-based photometry added
with model-derived ground-based constraints (as seen in
Figure~\ref{fig:zpdfs}).  The addition of ground-based constraints
perhaps makes the distinction between low-redshift and high-redshift
solutions difficult in \galg\ because of the relative offset in flux
between F150W and UltraVISTA $H$-band.  Despite the additional
ambiguity surrounding \galg, we keep it in the high-$z$ sample for
further analysis.

\subsection{Candidates at $z>13$}\label{sec:z13}

In an effort to explore the most extreme subset of new discoveries
within the EoR, we have also identified a sample of candidate galaxies
at $z_{\rm phot}>13$ from our existing imaging. As a natural
consequence of the design of our survey, sources with $z_{\rm
  phot}>12.5$ are ostensibly only detected in F277W and F444W.  None
are sufficiently bright to be detected with {\it Spitzer} at
3.6\,\um\ (a wavelength not covered with {\it JWST} imaging in
COSMOS-Web), UltraVISTA filters, or MIRI at 7.7\um.  Beyond the
limited filter set, the wavelength gap between F277W and F150W
presents another challenge making it difficult to quote photometric
redshifts more precise than $\Delta z\approx1$.  Another consequence
of the limited photometry is that the difference in $\chi^2_{n}$
between low- and high-redshift fits is diminished: $\chi^2_{n}$ is
overall lower because only two bands have ${\rm SNR}>3$.  Indeed, all
sources in this category fail at least one $\Delta\chi^2_{n}$ cut
(primarily {\tt EAzY} and {\tt LePhare}).

Our approach towards candidates in this regime is thus somewhat
conservative.  Of an initial set of 31 sources in our initial {\tt
  EAzY} catalog fit to photometric redshifts $z_{\rm phot}>12.5$ with
$[F277W]<27.5$, we down-select to five viable candidates in this
paper.  Sources are rejected from the sample because they fail all
$\Delta\chi^2_{n}>0.6$ criteria (for {\tt EAzY}, {\tt LePhare} and
{\tt Bagpipes}) and have $>$5\%\ probabilities of $z<7$ solutions,
they have either diffuse morphologies with radii $R_{\rm
  eff}>0\farcs5$ or they are spatially unresolved in F277W ($R_{\rm
  eff}<0\farcs15$), or they are sufficiently red ($[F277W]-[F444W]<0$)
to cast significant doubt on a $z>13$ solution.  As a natural
consequence of this approach, the five remaining sources---\gali,
\galj, \galk, \gall, and \galm---span somewhat bluer colors than the
parent population with $-0.7<[F277W]-[F444W]<-0.3$.  All have
integrated probabilities of being at $z<7$ less than 0.4\%\ using the
fiducial model-based photometry.

  We note that \gall\ and \galm\ both have MIRI coverage at
  7.7\um\ though neither is detected.  The measured flux densities we
  obtain for them using {\tt SE++} model-based photometry are $S_{\rm
    7.7\mu m}=26\pm31$\,nJy and $S_{\rm 7.7\mu m}=49\pm31$\,nJy,
  respectively.

Of the five candidates, only two show significantly elevated
low-redshift probabilities using {\tt SE} `classic' aperture
photometry: \gali\ has 11.9\%\ of the PDF at $z<7$ and \galm\ has
7.2\%\ at $z<7$.  When adding the aperture photometry with
ground-based constraints the $z<7$ probability is reduced in \galm\ to
1.8\%, but is much higher (98.4\%) for \gali.  Though this result is
inconsistent with our model-based constraints on \gali, we remove it
from future analysis in the discussion.  We also remove source
\galk\ from further analysis because it fails all $\Delta\chi^2_{n}$
criteria, even though the integral of its $z<7$ probability is
$<$0.5\%.

For the purposes of the discussion and ensuing physical
characteristics, we only retain three sources in the $z>13$ sample:
\galj, \gall, and \galm, but we provide descriptions of all five.  We
continue to stress that this sample is overall less robust than
sources described in \S~\ref{sec:superbright} and \S~\ref{sec:bright} and
all, including those we have thrown out on suspicion they are low-$z$,
require spectroscopic confirmation.

\section{Discussion}\label{sec:discussion}

Here we present a more detailed discussion of the ramification of
these discoveries.  First we present a discussion regarding
low-redshift interlopers, a direct measurement of
their volume density and contribution to the UVLF, and a summary of
their measured physical characteristics.  We then present a more
detailed discussion of their stellar mass estimates and their implied
star-formation efficiencies within $\Lambda$CDM.  We then discuss the
potential gas content of the systems and last raise the possibility of
future observations constraining the host galaxies' halo masses, which
could also inform constraints on alternate cosmological frameworks.

\subsection{The Possibility of Low Redshift Interlopers}\label{sec:lowz}

{\it JWST} observations of high
  redshift galaxies so far have made clear that galaxies at $z<7$
with strong emission line equivalent widths in the rest-frame optical
can masquerade as ultra high redshift ($z>10$) galaxy candidates in
JWST's broadband filters
\citep{zavala22a,mckinney23a,fujimoto22a,naidu22b}.  While discussion
of this phenomenon arose prominently via model fitting to CEERS-93316,
an exceptionally exciting and bright $z\sim16$ candidate
\citep{donnan23a,zavala22a,naidu22b}, recent NIRSpec follow-up
confirming a low redshift solution ($z\sim4.9$) highlights the
complexity and difficulty in selecting robust, ultra high-redshift
candidates \citep{arrabal-haro23b}.  This type of strong emission line
contaminant may only be a substantial concern for $z\sim4.6-4.8$
contaminants in fields where coverage exists in a majority of NIRCam
broadband filters.  At those redshifts, strong emission line sources
may appear as Lyman-break candidates at $z>15$.  However, COSMOS-Web
has fewer filters and so emission line contaminants are possible over
a broader range of redshifts.  Examples of best-fit low redshift
solutions (restricted to $z<7$) are shown in gray in
Figure~\ref{fig:bio1}.

Figure~\ref{fig:zpdfs} shows three different derivations of the full
redshift probability density distributions from $0<z<20$ for each
source measured using {\tt EAzY} assuming flat redshift priors.  We
compare our fiducial model-based photometric constraints to aperture
photometry.  In all cases, the probability of a $z<7$ solution using
model-based photometry is $<$3\%\ for each source.  The source with
the highest probability at $z<7$ is \gald\ with 2.4\%, likely caused
by its comparatively red color; the rest of the sample has
P($z<7$)$<$1\%.
We emphasize again that the {\tt SE++} model-based and
{\tt SE} classic aperture-based photometry are independently measured:
the first from images in their native resolution, and the second from
PSF-homogenized images. Because the aperture photometry only includes
measurements from five bands ({\it Hubble} F814W and the {\it JWST}
NIRCam bands), its derived redshift PDFs are broader and generally
show an increased probability for a low redshift solution.  This
increased uncertainty in the redshift PDF can be attributed to the
limited number of bands used to constrain the redshift.
To demonstrate the importance of the ground-based photometry for the
photometric redshifts, we construct a hybrid photometric catalog which
marries the {\it Hubble}+{\it JWST} constraints of the aperture
photometry to the model-derived ground-based observations.  We note
that for these $z>10$ candidates, the ground-based constraints are
almost all modestly constraining non-detections (or a handful of low
SNR detections in UltraVISTA).  As a result, this hybrid catalog can
be thought of as a sanity check on our model-based photometry, as we
have replaced the most constraining, high-SNR bands
(i.e. {\it JWST} NIRCam) with aperture photometric measurements. In
all but two cases, \galf\ and \gali, those additional ground-based
constraints significantly reduce the probability of a viable low-$z$
solution over the space-based only PDFs.  \galf\ is
kept in the sample given that it formally passes our $\chi^{2}_{n}$
criteria for inclusion and \gali\ removed as previously discussed in
\S~\ref{sec:z13}.

  An important caveat of the previous paragraph is the
  adoption of {\it flat} redshift priors which is a somewhat standard,
  though potentially flawed, literature convention.  The on-sky
  surface density of galaxies as a function of redshift declines
  steeply with increasing redshift, regardless of brightness. For
  example, the on-sky surface density of a $z\sim4$ 27$^{th}$
  magnitude (AB) source is $\sim$300--500 times higher than similarly
  bright $z\sim10$ sources according to some of the latest
  compilations of the UVLF \citep[e.g.][]{finkelstein16a,harikane23a}.
  By this argument, any source with a low redshift peak exceeding
  0.3\%\ may have as much of a 50\%\ chance of being a true low
  redshift source, and those with 3\%\ may be 10$\times$ more likely
  low-$z$ than high-$z$. Nevertheless, such a thought experiment does
  not adequately account for variations in the intrinsic SEDs, the
  lack of detection in deep optical stacks, or the measured sizes of
  the sample.  A dedicated study focused on best practices in
  photometric redshift fitting to such samples would be timely though
  beyond the scope of this work.

Given the broader range of potential contaminants in COSMOS-Web, we
explored if the ensuing derived parameters for such low-redshift
solutions in $M_\star-{\rm SFR}-A_{V}$ space were physical.  Digging
into the posterior distribution of physical properties from {\tt
  Bagpipes} low redshift fits, we find the redshift range of probable
contaminants spans $1<z<4$ with median redshift $\langle z\rangle=3$,
stellar mass $\langle M_{\star}\rangle=10^{9}$\,\msun, $\langle {\rm
  SFR}\rangle=10$\,\sfr, and attenuation $\langle A_{V}\rangle=1.7$.
The attenuation (thus reddening of stellar continuum), combined with
high star formation rates, is what is needed to reproduce the
photometry of a $z>10$ Lyman break.  Note that allowing $A_{V}$ to
vary up to 6 generates another cluster of possible solutions at $z<1$
with $A_{V}\sim5$: these are largely unphysical, as they would imply
extreme attenuation and high SFRs in low mass dwarf galaxies
\citep{bisigello23a}.  If such a phenomenon were common, submillimeter
number counts would likely be dominated by such sources, and they are
not \citep[\citealt*{casey14a};][]{fujimoto16a}.

Even with more reasonable limits set on $A_{V}<3$, we note that one
may expect a non-negligible fraction (20-30\%) of contaminants to be
detectable in existing (sub)millimeter imaging in the field; we
estimate this fraction by inferring $A_{UV}$ from $A_{V}$ (where
$A_{UV}\approx2.6 A_{V}$), converting to ${\rm IRX}\equiv L_{\rm
  IR}/L_{\rm UV}$, and thus scaling $M_{\rm UV}$ to $L_{\rm IR}$.
Sources above $\sim10^{12}$\,\lsun\ would be detectable in existing
SCUBA-2 and AzTEC maps in the field
\citep{aretxaga11a,casey14a,simpson19a}.  None of our sample are
detected above $3\sigma$ detection limits in those datasets.

\begin{figure}
  \centering
  \includegraphics[width=0.99\columnwidth]{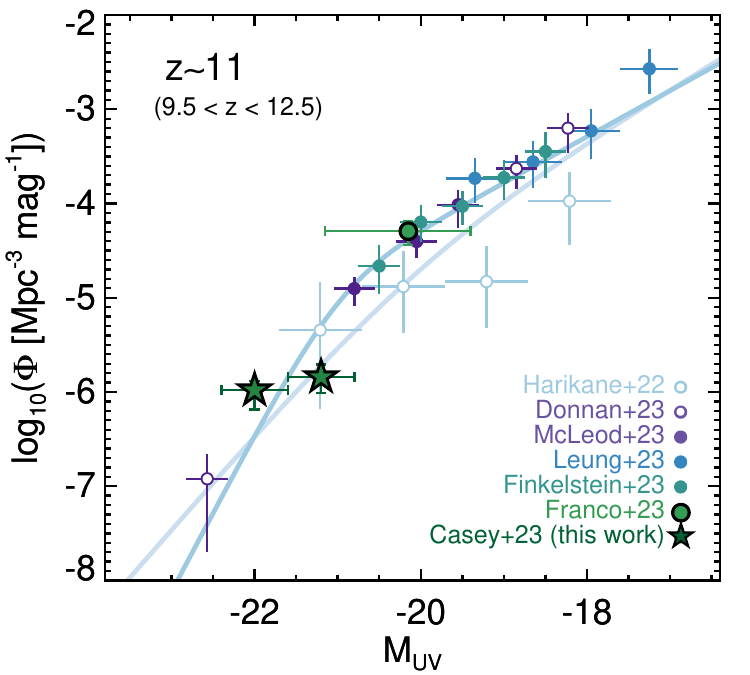}
  \caption{The direct contribution of sources presented in this paper
    to the UV luminosity function at $z\sim11$ (green stars) against
    recent {\it JWST} literature measurements \citep[][Finkelstein
      \etal, in
      preparation]{harikane23a,donnan23a,mcleod23a,leung23a,franco23a}.
    The two functional fits shown are the \citet{donnan23a} $z\sim10$
    double powerlaw fit (lighter blue) and the \citet{leung23a}
    $z\sim11$ double powerlaw fit (darker blue).  The tabulated UVLF
    data from this paper are provided in Table~\ref{tab:uvlf} along
    with a measurement at $z=14$ (not shown here).  The $z\sim10$
    measurement from the first epoch of COSMOS-Web data in
    \citet{franco23a} is shown in light green.  These measurements do
    not account for incompleteness, but do account for contamination
    by adjusting the contribution of each source by the probability
    that it is indeed $z\simgt10$.  Our measurements are in agreement
    with other literature estimates of the UVLF at similar redshifts.}
  \label{fig:uvlf}
\end{figure}

\subsection{Volume Density and UVLF Contribution}

The solid angle covered by COSMOS-Web to-date is 0.28\,deg$^2$,
implying the survey volume between $9.5<z<12.5$ of
  $\sim$4.5$\times10^{6}$\,Mpc$^{3}$.
  This redshift
    interval brackets the more confident candidates that are detected
    in more than two bands and the average redshift for galaxies in
    the bin is near $z\sim11$.  While our $z>13$ candidates have
  redshift PDFs extending out to $z\sim20$, solutions at $z>15$ are
  unlikely, so we cap the volume estimate relevant for those sources
  at $\sim$2.4$\times10^{6}$\,Mpc$^3$ corresponding to
    $13<z<15$ (the $z\sim14$ sample).

Figure~\ref{fig:uvlf} shows the direct contribution of detected
sources (calculated using the 1/$V_{\rm max}$ method)
presented in this paper to the UV luminosity function at 
  $z\sim11$,
  with  measurements  also provided in
Table~\ref{tab:uvlf}. We calculate the contribution of
  these sources to the UVLF through 500 Monte Carlo draws from the
  posterior distributions of measured \muv\ and redshift from {\tt
    Bagpipes}, where sources may be counted in different magnitude
  bins in different draws or fall outside of the designated redshift
  range. These measurements have not been corrected for
incompleteness, yet they do address potential
contamination. Broadly, we find that the volume density of very bright
galaxies found in COSMOS-Web is  well aligned with
expectation from  other recent {\it
    JWST}-measured luminosity functions at $z>10$. Figure~\ref{fig:uvlf} shows
  measurements at $z\sim10-12$ from \citet{donnan23a},
  \citet{harikane23a}, \citet{mcleod23a}, \citet{leung23a},
  \citet{franco23a} and data from the CEERS collaboration
  (S. Finkelstein, private communication).  The two functional double
  powerlaw fits shown are from \citet{donnan23a} and
  \citet{leung23a}.

    Most noticeable from our data is the relative flat slope
  of the UVLF at the bright end; this could be caused by
  incompleteness in our lower luminosity bin, which we do not correct
  in this work.  Considering that the magnitude bin at \muv$=-22$ is
  relatively complete, our data disfavors a Schechter
  function fit to the UVLF similar to other work.  A more thorough
observational derivation of the UVLF from COSMOS-Web will follow in a
forthcoming paper (Franco \etal, in prep) and it will include
completeness simulations and an extended measurement down to the
threshold detection limit of the survey.

\input{tab_uvlf}

\subsection{Physical Properties of Bright $z>10$ Candidates}

\begin{figure*}
  \centering
\includegraphics[width=0.99\columnwidth]{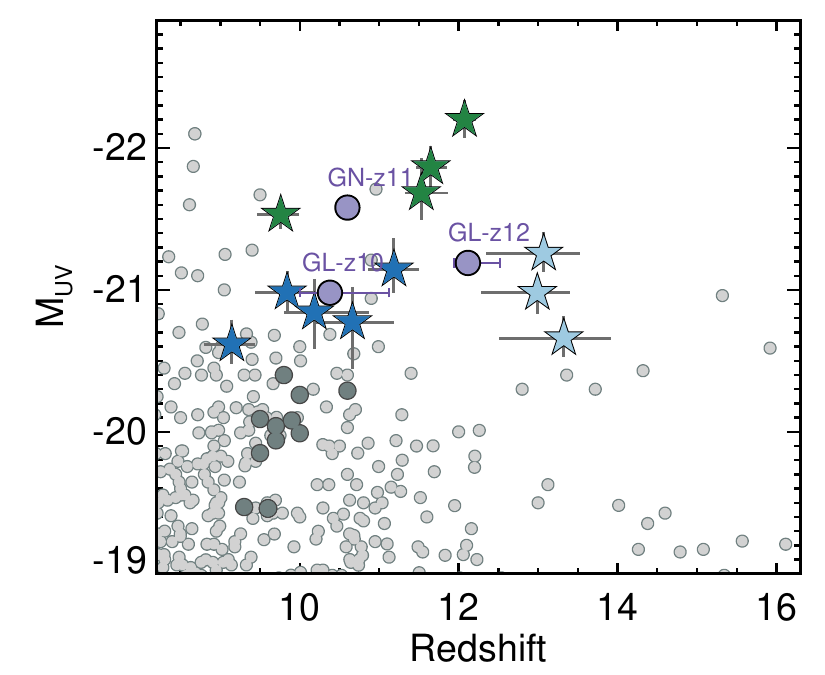}
\includegraphics[width=0.99\columnwidth]{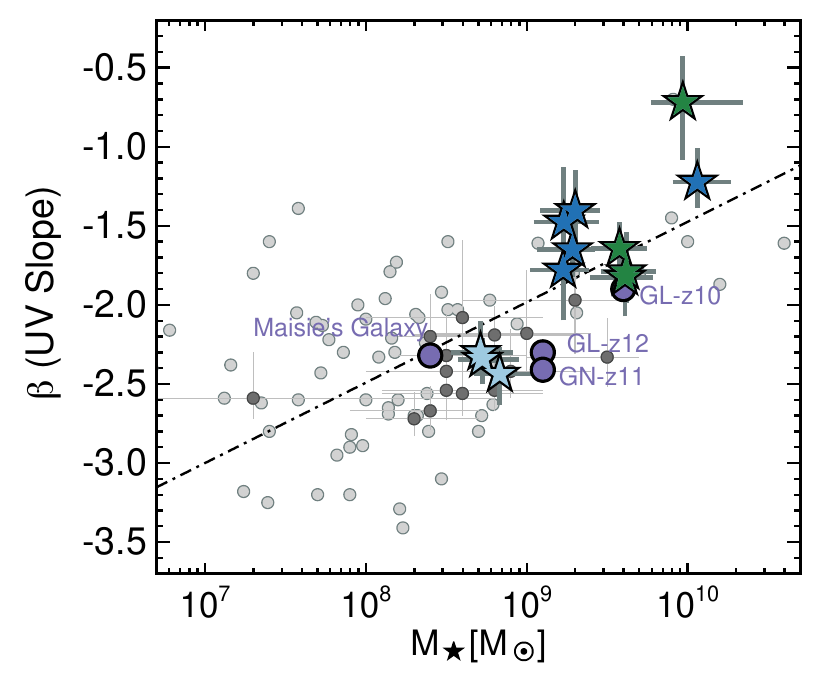}
\caption{The derived physical characteristics of our luminous $z>10$
  candidates relative to other samples of early Universe galaxies in
  the literature (gray points, the majority of which are from the
  JADES sample in the left panel, \citet{hainline23a}), explicitly
  highlighting the COSMOS-Web sample found by \citet{franco23a} in
  dark gray.  The four galaxies that are exceptionally luminous are
  shown as green stars, the bright $10\le z\le12$ sample shown as dark
  blue stars, and the $z>13$ sample as light blue stars; this color
  scheme persists in Figures~\ref{fig:size}, \ref{fig:mstar}, and
  \ref{fig:sfh}.
  {\it Left:} The rest-frame absolute UV magnitude against redshift in
  comparison to literature samples. We highlight three other luminous
  $10<z<12$ sources: GN-z11 \citep[now spectroscopically confirmed at
    $z=10.60$;][]{oesch16a,bunker23a}, GL-z10 and GL-z12
  \citep{castellano22a,naidu22a} with tentative spectroscopic identifications from
  ALMA \citep{bakx23a,yoon23a}.
  {\it Right:} The rest-frame UV slope $\beta$ with redshift.  Our
  sample is the reddest subset of candidates reported in the
  literature \citep{finkelstein12a,tacchella22a}.
  The dot-dashed line is the best-fit relation derived for $z=8$
  galaxies from \citet{finkelstein12a}.  No trend is detected in
  $M_{\rm UV}$ versus $\beta$.
}
\label{fig:phys}
\end{figure*}

We show in Figure~\ref{fig:phys} the distribution of sources presented
in this paper in absolute UV magnitude, rest-frame UV slope ($\beta$),
and stellar mass.  We compare to other reported candidates in the
literature summarized recently in \citet{franco23a}.
The four brightest sources discussed in \S~\ref{sec:superbright} are
shown in green in each panel.  Those four sources have luminosities
well matched to and exceeding GN-z11 at similar redshifts.  Given the
wide area covered by COSMOS-Web to-date, it is clear that we are
sensitive to the discovery of more intrinsically luminous sources
brighter than $M_{\rm UV}\approx-20.5$ beyond $z\sim10$.  At fixed
redshift, most {\it JWST} discoveries from deeper but narrower fields are
about 1-3 magnitudes (or $2.5-15\times$) fainter than the sample
presented herein.

The rest-frame UV slopes of this sample are a bit redder than most
Lyman-break galaxies (LBGs) at this epoch (the median presented in
  the literature at $z>8$ is $\langle\beta\rangle=-2.3\pm0.5$ and the
  median of this sample is $\langle\beta\rangle=-1.7\pm0.5$); one
source, \gald, is a significant outlier with $\beta=-0.6$ while all
others are bluer than $\beta=-1.2$.  Our sample is likely more red
than other samples for a few reasons: as exceptionally bright sources,
they tend to have higher estimated stellar masses \citep[see earlier
  measurements of this relation from][]{finkelstein12a,tacchella22a}.
  Those works measured direct correlation between $\beta$
  and $M_\star$ but not $\beta$ and \muv. With higher masses, it is
more likely that the stellar population is generally older or the
star-formation history more complex, such that there are either
proportionally fewer O stars contributing to the rest-frame UV flux or
the dust reservoirs in such early galaxies has built up enough to
redden the UV a small amount \citep[cf.][]{ferrara23a,ziparo23a}.
While more enhanced metallicity may also account for relatively red
$\beta$ slopes compared to low-metallicity galaxies with similar
star-formation histories, $\beta$ above $-2.0$ points to either dust
attenuation or less recent star-formation as the cause of the flatter UV slope.
  We do issue some caution in the interpretation of the
  $M_\star-\beta$ relationship, as both quantities are derived from
  SED fitting and have non-negligible covariance.

While bluer UV slopes are found
  in our highest redshift ($z>13$) sample, we caution that this is likely a selection effect:
galaxies with redder values of $\beta$ at $z\sim13$ would have
significant degeneracies with low-redshift solutions and thus be
removed from our sample for failing our selection criteria.  This
bias, which is prevalent in selection of LBGs at all redshifts, cannot
easily be addressed without significant investments in spectroscopy
for large samples.

\begin{figure}
  \includegraphics[width=0.99\columnwidth]{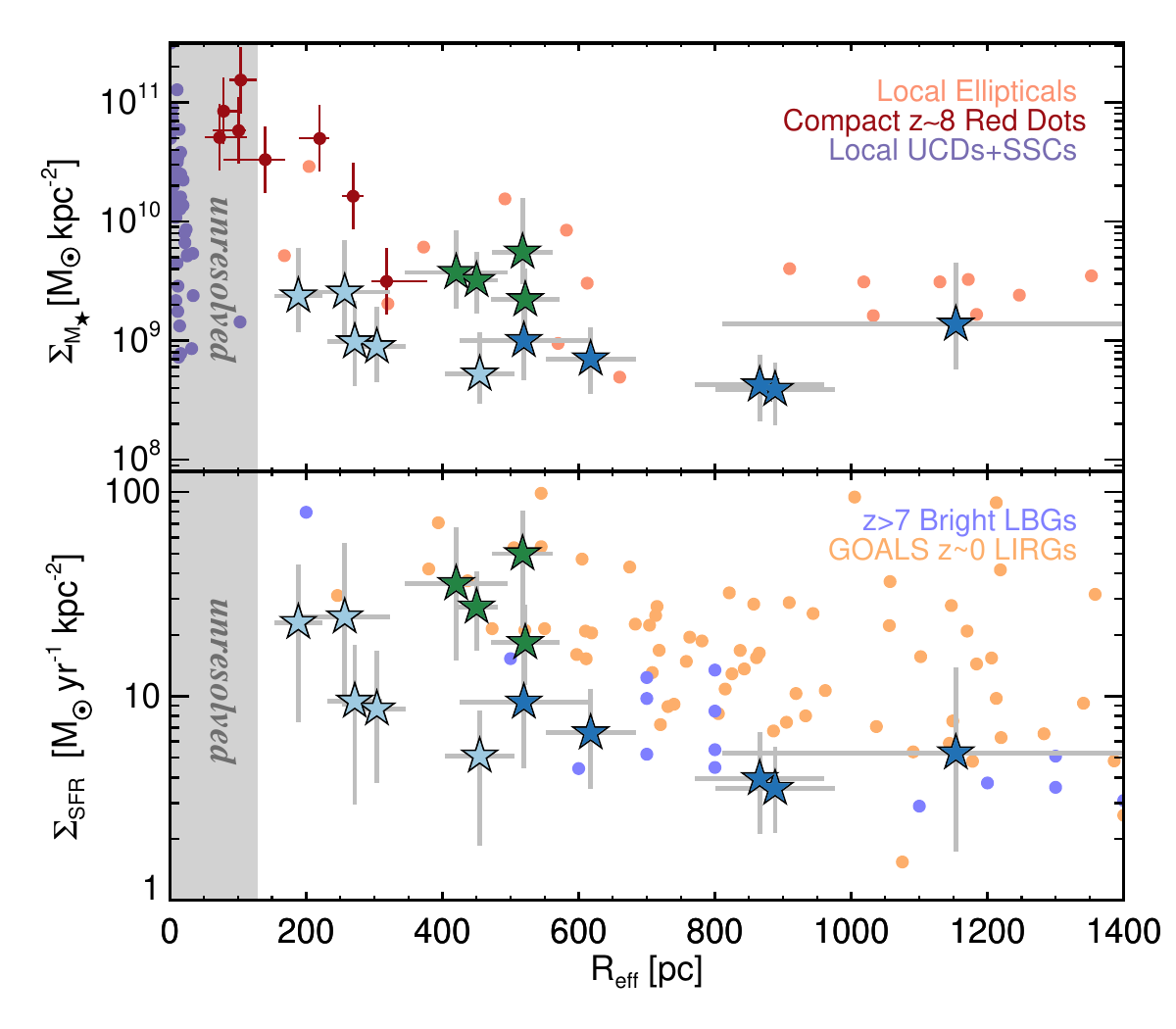}
  \caption{The F277W-measured half light radii of the sample plotted against the
    stellar mass surface density (top panel) and star formation
    surface density (bottom panel).  At top, we overplot the average
    stellar mass surface densities of local elliptical galaxies from
    \citet{lauer07a} and Ultra Compact Dwarfs and Super Star Clusters
    from the compilation in \citet{hopkins10a}.  In addition, we
    overplot the measured sizes and surface densities of the compact
    $z\sim8$ sources first discovered by \citet{labbe22a} and whose
    sizes were analyzed in \citet{baggen23a}.  The star formation
    surface densities are on par with other $z>7$ bright LBGs
    \citep{bowler17b}, local starbursts \citep{armus09a,mckinney23b},
    and submillimeter galaxies with measured sizes \citep[with $R_{\rm
        eff}>$1\,kpc;][]{burnham21a}. Sources from our sample are
    colored by subsample as in Figure~\ref{fig:phys}.}
\label{fig:size}
\end{figure}

Figure~\ref{fig:size} shows the sizes of galaxies in the sample
against stellar mass surface density and star formation rate surface
density.  All sources are spatially resolved with average sizes
$\langle$R$_{\rm eff}\rangle\,\sim\,$500\,pc, which reduces concern
that their emission may be dominated by AGN (though an unresolved
morphology would not guarantee it).  Stellar mass and star-formation
rate surface densities are calculated by dividing the total $M_\star$
or SFR by two to account for the $M_\star$ or SFR internal to $R_{\rm
  eff}$, then we divide by $\pi R_{\rm eff}^2$.  The stellar mass
surface densities are very similar to some of the most compact local
elliptical galaxies \citep{lauer07a,hopkins10a}, and Ultra Compact
Dwarfs and Super Star Clusters
\citep{hasegan05a,evstigneeva07a,hilker07a,mccrady07a} though our
sample is about 10 times larger in R$_{\rm eff}$.  This may suggest
that, as observed, these galaxies could be viable progenitors of
elliptical galaxies' cores with similar densities.  The star formation
surface densities are very similar to those seen in some $z>7$ bright
Lyman-break Galaxies \citep[e.g.][]{bowler17b}, some local starbursts
\citep[e.g. the Great Observatories All sky LIRG Survey, GOALS,
  sample,][]{armus09a,mckinney23b} and high redshift submillimeter
galaxies \citep{burnham21a,hodge16a}, though the latter systems tend
to be much larger ($R_{\rm eff}>1$\,kpc) with higher star formation
rates.

\subsection{Stellar Mass Uncertainties}\label{sec:stellarmass}

Mass derivations from rest-frame UV data are naturally uncertain.
Nevertheless, {\it JWST} provides a longer wavelength lever arm than
{\it Hubble}, into the rest-frame optical to constrain the SEDs of
galaxies even beyond $z>10$.  While the few bands and lack of coverage
in the rest-frame NIR misses the gold standard of stellar mass
derivation, the young age of the Universe at $z>10$ ($<$500\,Myr)
significantly narrows the dynamic range of possible star-formation
histories of luminous LBGs, and thus places reasonable limits on the
mass-to-light ratio, thus the underlying stellar mass of individual
sources.  Given the potential implications of stellar masses exceeding
$10^{9}$\,\msun\ at $z>10$ (all of our sample exceeds these limits),
we address possible sources of uncertainty in our stellar mass
derivation here.

We first check the sensitivity of the adopted star formation history (SFH)
on the resultant stellar mass; generally, star formation that occurs
further in the past will have a higher mass-to-light ratio implied for
the rest-frame UV, thus higher stellar mass.  We compare our fiducial
{\tt Bagpipes} model, which superimposes a delayed-$\tau$\ SFH with a
recent, constant starburst to a model with {\it only} a delayed-$\tau$
SFH.  The delayed-$\tau$-only SFH models
effectively increase the stellar mass estimates for
  fixed photometry by factors of 0.4--3$\times$ \citep[see
  also][]{michaowski12a,michaowski14a,mitchell13a}.  Conversely, we can
ask what fraction of stellar mass in our sample assembles during the
recent, constant starburst phase; in a majority of cases,
$\simgt$99\%\ of the stellar mass is attributed to a recent burst
(forming within the previous 50\,Myr on average).  If we
allow for an even more extreme and recent burst without
  the contribution from the delayed-$\tau$ model, the stellar masses are only
reduced by $\sim$10\%\ below the fiducial estimates, well within reported uncertainties.
In that sense, this illustrates that the stellar masses we derive in
this work are a conservative lower limit for a normally-behaved
stellar population: one comprised of population II and I stars with
low, but not extremely low, metallicity.

If population III stars dominate the light in the rest-frame UV, their
expected top-heavy IMF \citep{hirano14a,hirano15a} would result in a
UV continuum dominated by nebular, rather than stellar, emission.
This would lead to a higher light-to-mass ratio from the UV
\citep[e.g.][]{schaerer09a} by factors of $\sim$0.5--0.6\,dex,
reducing the highest mass estimates in our sample at $z\sim12$ from
$\sim5\times10^{9}$\,\msun\ to $\sim10^{9}$\,\msun.  This would, of
course, imply that first generation star formation dominates the
energy output of these systems, which may be a difficult and somewhat
extreme boundary condition, even for such high redshift galaxies given
that their estimated halo masses are quite high, $M_{\rm
  halo}\sim10^{11}$\,\msun.  Another likely consequence of the
rest-frame UV being dominated by population III stars would be a much
bluer slope, $\beta$, than measured here.

Lastly, we consider the impact of AGN on our stellar mass estimates.
It has become clear that actively accreting supermassive black holes
may be a fairly prominent source of energy output at early times, and
{\it JWST} has enabled the identification of AGN with lower mass black holes
out to earlier times \citep{larson23a}.  For AGN to significantly
impact the mass in our sample, they would need to dominate the
rest-frame UV luminosity by a factor of several over the stellar
contribution.  This would effectively translate to a lower limit on
AGN luminosity of $M_{\rm UV}<-20.5$ or $L_{\rm
  UV}\simgt1-2\times10^{10}$\,\lsun.  At an Eddington ratio of
$\epsilon\sim0.1$, this would imply a minimum black hole mass of
$M_{\rm BH}\sim5-6\times10^{6}$\,\msun.  Such masses would be somewhat
unexpectedly large for the downward-revised stellar mass estimates of
their host galaxies, $\simlt10^{9}$\,\msun, so we assess this outcome
as less likely than our stellar mass estimates without significant AGN
contribution.   While some high-$z$ supermassive black
  holes do seem unusually massive for their host galaxies
  \citep{kocevski23a,larson23a}, it does not appear that they
  contribute significant emission to the continuum.

\begin{figure}
  \centering \includegraphics[width=0.99\columnwidth]{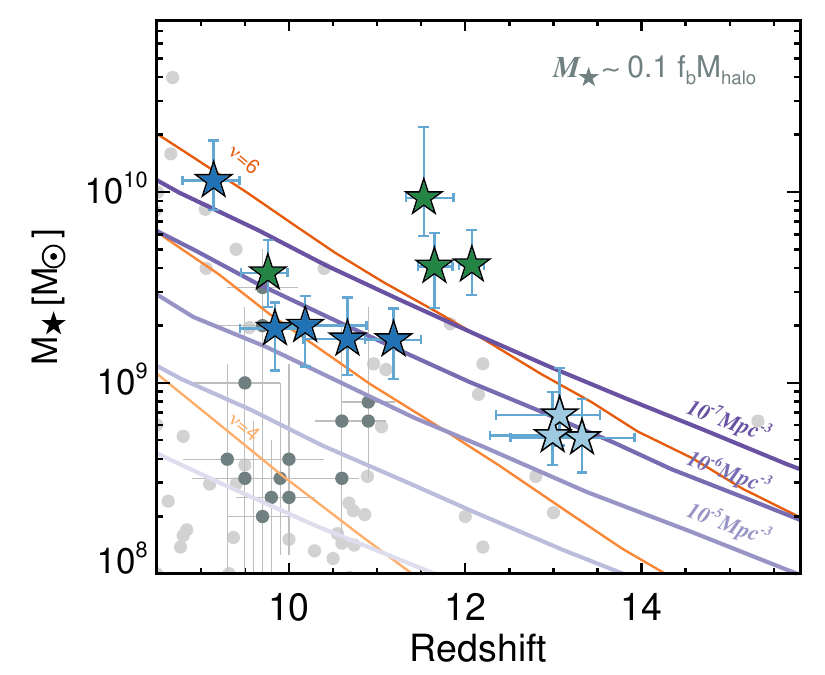}
  \caption{Stellar mass versus redshift for our sample relative to
    curves of fixed volume density, peak height, and other candidate
    galaxies reported in the literature (light gray points).  Volume
    density curves (purple) and peak height (orange) are generated
    from expectation of the halo mass function scaled by the cosmic
    baryon fraction $f_{\rm b}$ and an efficiency to convert baryons
    to stars of $\epsilon_{\star}=0.1$ as in
    \citet{boylan-kolchin23a}.  The volume in the current COSMOS-Web
    dataset is a $\sim$few $\times10^{6}$\,Mpc$^3$ in a bin of width
    $\Delta z=2$. This comparison highlights the particularly unusual
    nature of the three massive galaxies at $z\sim12$, \galb, \galc,
    and \gald, whose existence defies expectation. Their presence
    demands either higher abundance of massive halos at $z\sim12$ or
    an enhanced stellar baryon fraction $\epsilon_{\star}\sim0.2-0.5$,
    implying that the star forming efficiency be elevated
    ($\epsilon_{\rm SF}\sim1$) for a significant fraction of the
    galaxies' star formation histories. The possible star formation
    histories of these systems is shown in Figure~\ref{fig:sfh} and
    more details on the implications of their masses explored in the
    discussion.  Sources from our sample are colored by subsample as
    in Figure~\ref{fig:phys}.}
  \label{fig:mstar}
\end{figure}

The stellar masses of our sample, as estimated using our fiducial {\tt
  Bagpipes} burst+delayed-$\tau$\ model, are shown against redshift in
Figure~\ref{fig:mstar}.

\subsection{Massive Beacons: the assembly of the first megalithic halos?}

\begin{figure}
  \includegraphics[width=0.99\columnwidth]{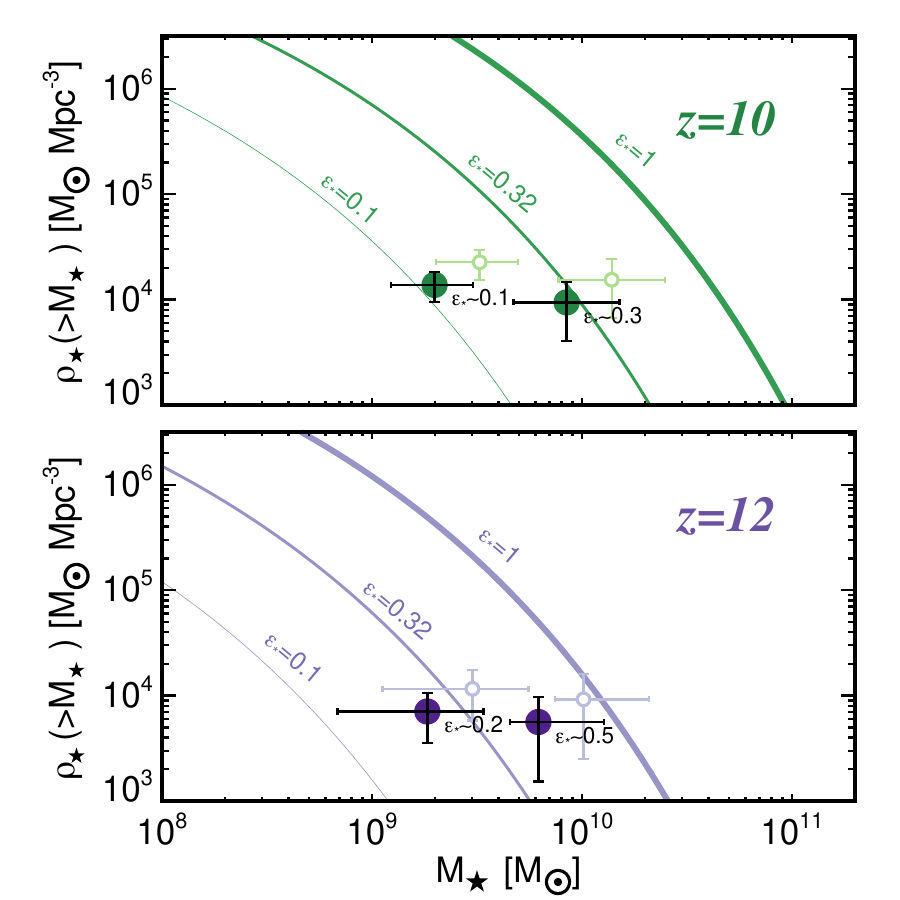}
  \caption{The cumulative stellar mass volume density as a function of
    stellar mass calculated at $z=10$ and $z=12$ from our sample; halo
    mass curves, with three different integrated stellar efficiencies
    or stellar baryon fractions: $\epsilon_{\star}=0.1$, 0.32 and 1,
    are derived using the methodology outlined in
    \citet{boylan-kolchin23a}.  Points are derived using the full
    posterior distributions in $z$ and $M_\star$ for each source in
    bins of width $\Delta z=1$ centered on either redshift.  At $z=10$
    we infer stellar baryon fractions $\epsilon_{\star}\sim0.1-0.3$,
    while at $z=12$ we infer higher stellar baryon fractions,
    $\epsilon_{\star}\sim0.2-0.5$. Open circles represent the
    extrapolated sum of stars and molecular gas; gas masses are
    inferred from the Kennicutt-Schmidt relation.  At the highest
    masses at $z=12$, it is apparent that all baryons could be
    comprised of stars and molecular gas, leaving little room for
    substantial reservoirs of, e.g., atomic gas.  This highlights the
    need to gather cold gas observations of the sample.}
\label{fig:mbk}
\end{figure}

In stellar mass space, a subset of our sample push the bounds of the
most massive sources that could plausibly be found in deep {\it JWST}
surveys at $z>10$: in particular, \galb, \galc, and \gald\ with masses
$M_\star\sim4-10\times10^{9}\,$\msun\ at $z\approx12$.  We overplot
curves of constant number density on Figure~\ref{fig:mstar} implied
from the halo mass function \citep{sheth99a} scaled by the cosmic
baryon fraction and a reasonable
`efficiency' of $\epsilon_{\star}\approx0.1$, where $\epsilon_{\star}$
represents the fraction of baryons that have been converted into stars
within a halo over its integrated lifetime, $\epsilon_{\star}\equiv
M_{\star}/(f_{b}M_{\rm halo})$.  We note that the typical
  peak of the stellar-mass-to-halo mass relation (SMHR) is
  $M_\star$/$M_{\rm halo}\sim1-3\times10^{-2}$ across a range of
  redshifts
  \citep{shankar06a,mandelbaum06a,conroy09a,behroozi10a,behroozi19a,shuntov22a},
  implying maximum $\epsilon_\star\sim0.2$ (and one might expect this
  fraction to be lower at much earlier cosmic times where constraints
  on the SMHR do not yet exist).
We also overplot curves of constant peak height $\nu$, a measure of
the fraction of mass contained in halos above a given mass threshold;
higher values of $\nu$ mark increasingly massive halos where $\nu=4.5$
  at $z=0$ corresponds to a halo of mass
  $\approx5\times10^{15}$\,\msun.  This makes clear
  that the implied evolution of our sample is extreme with $\nu>6$
  given $\epsilon_{\star}=0.1$: they represent the highest possible
  mass overdensities that will grow to host massive galaxy clusters in
  the present-day Universe.

In Figure~\ref{fig:mbk}, we directly calculate the cumulative stellar
mass density in two bins centered at $z=10$ and $z=12$ with width
$\Delta z = 2$.  We use the full posterior distributions in $z$ and
$M_\star$ as derived using {\tt Bagpipes} for these calculations and
assume Poisson uncertainties.  Direct comparison to the halo mass
function implies reasonable to high stellar baryon fractions with
$\epsilon_{\star}\approx0.1-0.3$ at $z=10$, not too distinct from
expectation from the stellar-to-halo-mass relation
\citep{behroozi19a}.  However, at $z=12$, our discoveries imply higher
stellar mass fractions, with $\epsilon_{\star}\sim0.2-0.5$ (see also
forthcoming publications by K. Chworowsky \etal\ and M. Shuntov
\etal).  \citet{harikane23a} present a comprehensive overview of early
Universe discoveries from {\it JWST} in its first year, and they find
that particularly bright galaxies ($M_{\rm UV}<-19.5$) found in
deeper, smaller volume surveys require high efficiencies\footnote{Note
that efficiency, $\epsilon$, in \citet{harikane23a} captures the
instantaneous `efficiency' of a halo whereas $\epsilon_\star$ is the
integrated efficiency; both are distinct from $\epsilon_{\rm SF}$, the
star-forming efficiency which is the inverse of the gas depletion
time, discussed more in \S~\ref{sec:gas}.}, where $\epsilon={\rm
  \dot{M_\star}}/(f_{b}\dot{M_{\rm halo}})\sim0.3$, to produce stellar
masses $\sim10^{8-9}$\,\msun\ at $z=12-16$.  Such candidates were not
expected before {\it JWST}.  Are such high stellar baryon fractions of
$\epsilon_{\star}\sim0.2-0.5$ realistic?

Some theoretical work suggests they are.  In particular, galaxies at
$z>10$, largely embedded in the neutral Universe before reionization,
would not be bombarded with a background of UV radiation and thus the
rapid collapse of molecular clouds could see very high rates of star
formation \citep{susa04a}.  The Feedback-Free Starburst (FFB)
presented in \citet{dekel23a} provides a detailed account
  from first principles of how such a starburst might be
powered; in such systems, the free-fall time is $<$1\,Myr and rapid
star formation occurs before massive stars develop winds and
supernovae feedback and the external UV background has not yet been
established. This is very similar to prior simulation
  work which has demonstrated certain regimes where feedback fails to
  regulate star-formation \citep{torrey17a,grudic18a}.

With a delayed onset of feedback, one
might expect  the
  instantaneous star-forming efficiency ($\epsilon={\rm
    \dot{M_\star}/(f_{b}\dot{M_{\rm halo}})}\approx1$) for short
periods of time ($\simlt$5\,Myr), leading to appreciably
  larger $\epsilon_\star$ of order a few tenths.  Galaxies residing in
  halos of mass $\sim10^{11}$\,\msun\ at $z\sim10$ may be expected to
  have FFB gas densities, thus they might have stellar masses as high
  as $\sim10^{10}$\,\msun, SFRs in the 10's of \sfr, and blue, compact
  (sub-kpc) morphologies.  This describes the properties of the
  $z\sim12$ massive subsample well: $\langle
  M_{\star}\rangle\approx5\times10^{9}$\,\msun, $\langle{\rm
    SFR}\rangle\approx40$\,\sfr, and $R_{\rm
    eff}\approx500$\,pc.  Future spectroscopy of such
    targets may further clarify applicability of the FFB model to such
    systems, particularly in measurements of metallicity and
    rest-frame UV slope (thus presence of dust).

      Another theoretical interpretation of the very luminous, early
      systems is provided in \citet{ferrara23a}, which suggests that
      short-lived bursts of super-Eddington star formation may blow
      out the majority of dust in early galaxies (with
      sSFR\,$>$\,20\,Gyr$^{-1}$), making it possible to detect very
      blue, luminous galaxies beyond $z>10$.  Such systems may have
      relatively high stellar masses $M_\star\sim10^{8-9}$\,\msun,
      metallicity $Z\approx0.1\,Z_\odot$, and blue rest-frame UV
      slopes ($\beta\sim-2.6$). \citet{ziparo23a} describes that
      either dust ejection by radiation pressure could result in such
      blue slopes, or alternatively, a patchy ISM with spatially
      distinct regions of obscured and unobscured stellar light.  Our
      sample is not quite as blue (with median
      $\langle\beta\rangle=-1.7\pm0.5$) as their fiducial model, which
      demonstrates some inconsistency with the super-Eddington dust
      ejection hypothesis, but the patchy attenuation model may indeed
      be applicable to these systems.  The median estimated specific
      SFR in our sample calculated on a 100\,Myr (10\,Myr) timescale
      is $\langle{\rm sSFR_{100}}\rangle=10^{+9}_{-4}$\,Gyr$^{-1}$
      ($\langle{\rm sSFR_{10}}\rangle=15^{+12}_{-8}$\,Gyr$^{-1}$).
      Future ALMA observations will provide crucial information as to
      the dust content of such systems.

\begin{figure*}
  \centering \includegraphics[width=0.9\textwidth]{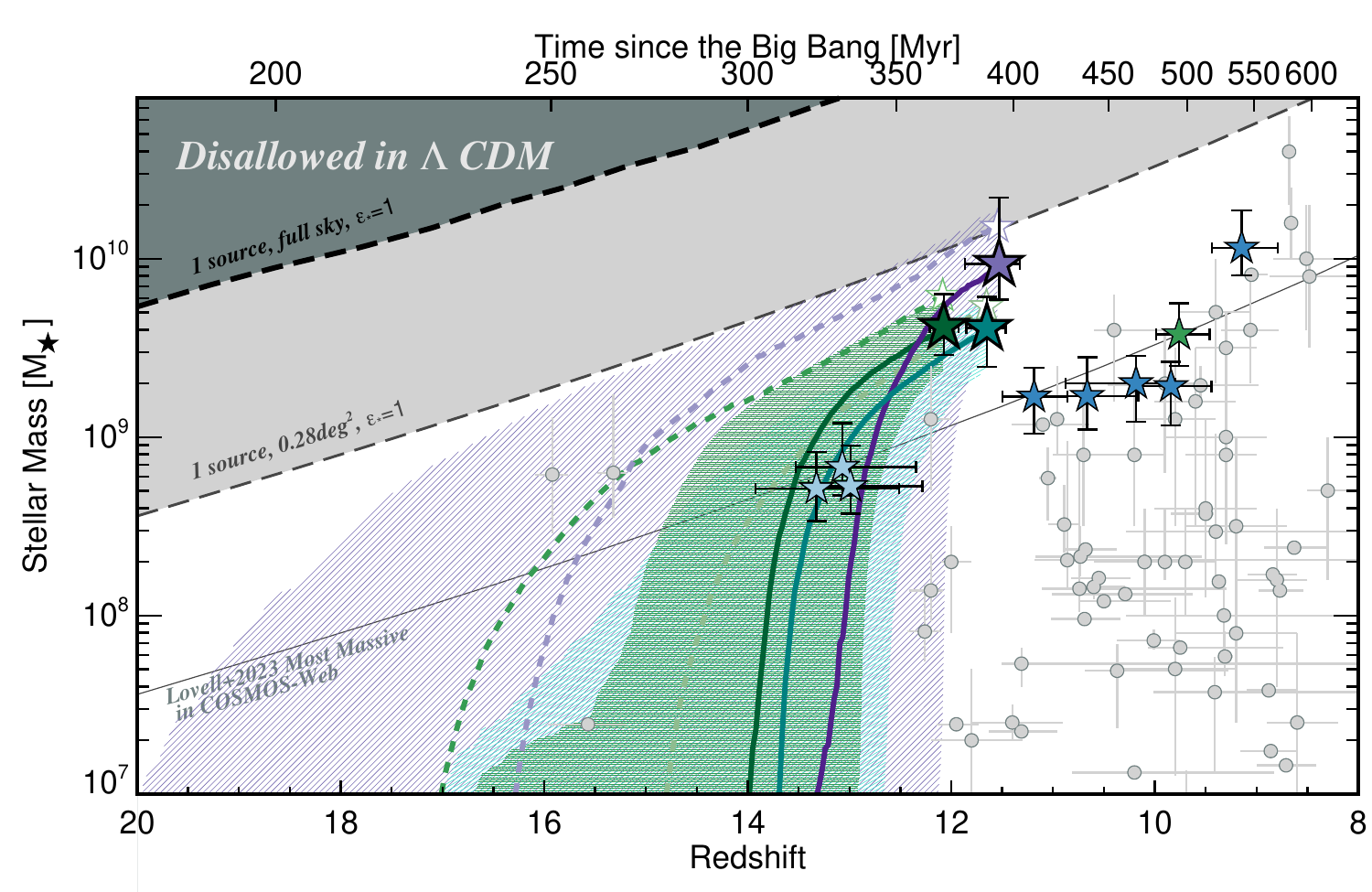}
  \caption{Stellar mass against redshift for the candidates identified
    in this paper (stars).  Here we highlight the integrated star
    formation histories (SFH) for the three most massive $z\sim12$
    candidates which incorporate a delayed-$\tau$ plus recent
    starbursts: \galb\ (dark green), \galc\ (teal), and
    \gald\ (purple). The dashed lines show stellar mass growth
    histories if the SFH is instead assumed to be delayed-$\tau$-only
    without a burst; this results in higher stellar masses that would
    have built up more mass early ($z>16$).  With the delayed-$\tau$
    plus burst model, the stellar masses may have increased by an
    order of magnitude in less than 100\,Myr, from $z\sim14$
    progenitors that look much like the candidates we identify in
    \S~\ref{sec:z13}.  Other high-$z$ candidate galaxies 
      from the literature are shown in gray points.  Stellar masses
    that are formally disallowed in $\Lambda$CDM are noted
    in dark gray, corresponding to the stellar mass
      threshold for the most massive halo in the full sky assuming
      $\epsilon_\star=1$. We also show the same threshold
      corresponding to 0.28\,deg$^2$, the area covered by COSMOS-Web
      in this work. Similarly, the
    most massive galaxy anticipated in all of COSMOS-Web, calculated
    by \citet{lovell23a}, is shown in the thin gray line; it assumes
    halo baryon-to-stellar conversion efficiency varies with average
    $\langle\epsilon_{\star}\rangle\sim0.06$,  and the
      confidence intervals about that mass threshold encompass the
      most massive $z\sim12$ galaxy (\gald) within $\sim$2$\sigma$.}
  \label{fig:sfh}
\end{figure*}

Figure~\ref{fig:sfh} shows another rendering of the stellar masses in
our sample against redshift. Here we have directly shown what the
progenitor population of the three most massive systems at $z\sim12$
might look like, via posteriors (inner 68\%\ confidence interval) on
the star formation histories, or cumulative stellar mass growth.  The
stellar mass growth in these systems are overwhelmingly dominated by a
recent burst, such that their stellar masses have grown at rates that
far outpace the growth of their parent dark matter halos (at a fixed
volume density).  As pointed out in \S~\ref{sec:stellarmass}, such
rapid and recent stellar mass growth provides the most conservative
stellar mass estimates for the sample as a whole.  The
  suggested burst-driven nature of \galb, \galc, and \gald\ in
  particular would imply that their host galaxy halo masses may
  intrinsically be lower than one might expect given their
  stellar masses, on par with some of the less luminous galaxies in
  our sample.  Recent work from the FIRE simulation \citep{sun23a} has
  shown that, indeed, no special adjustments are needed to reproduce
  the observed characteristics of very bright early {\it JWST}
  discoveries, which they find are driven by stochastic and recent
  bursts of star formation.

    We should note that in most, if not all, theoretical
  models built to explain very massive $z>10$ galaxies, rapid bursts
  of star formation happen on {\it short} timescales, $\simlt$5\,Myr.
  Our SED fitting using {\tt Bagpipes} allows bursts to have either a
  very short or relatively long duration, up to 100\,Myr.
  Unfortunately limitations in the current dataset do not enable
  meaningful direct constraints on burst timescale (i.e. the posterior
  distribution in ages of the burst component is flat). However, it
  may be possible that such exceptionally luminous systems have
  experienced a series of short bursts that are fairly well modeled by
  one long-duration burst up to 100\,Myr.

The ascent of such massive $z\sim12$ systems is so rapid that the
population we identify at $z>13$ --- with stellar masses an order of
magnitude lower --- could plausibly serve as a progenitor population,
despite the short timescale ($<$70\,Myr) between the two epochs.  At
later times, it is possible galaxies like \galb, \galc, and
\gald\ evolve to become some of the universe's first massive galaxies
\citep{carnall23a,glazebrook23a}.

\subsection{Not all baryons are stars}\label{sec:gas}

Most baryons in galaxy halos beyond $z>3$ should be contained in gas and
not stars \citep{walter20a}.  What ramifications do such high stellar
masses in our sample have on the potential to observe their reservoirs
of molecular and atomic gas?  Such observations may yet prove crucial
to our interpretation of their masses, thus efficiencies.

First, it is worth recognizing that typical observed efficiencies in
the star formation process rarely exceed $\epsilon_{\rm SF}\sim0.1$
\citep{evans09a,bigiel10a,kennicutt12a}.  In the context of galaxies'
gas supply, star-forming efficiency is defined $\epsilon_{\rm
  SF}\equiv {\rm SFR}/M_{\rm gas}$ (the inverse of the gas depletion
timescale) and normalized to 10$^8$\,yr, thus represents the fraction
of gas consumed every 100\,Myr.  This is {\it not} the same as
$\epsilon_{\star}$, which we call a stellar baryon fraction in this
work, but others refer to as star-formation efficiency;
$\epsilon_\star$ may be thought of the integral form of $\epsilon_{\rm
  SF}$.  Similarly $\epsilon\equiv{\rm \dot{M_\star}/(f_{b}\dot{M_{\rm
      halo}})}$ is also not the same as $\epsilon_{\rm SF}$, as the
later captures only baryonic processes.  If we approximate $M_{\rm
  gas}\approx f_{b}M_{\rm halo}$ (which is a firm upper limit to
$M_{\rm gas}$) with SFR$\approx50$\,\sfr, we find the average
star-forming efficiency would need to be $\epsilon_{\rm SF}\ge0.32$,
in excess of limits seen in local molecular clouds but necessary to
build the observed stellar populations.

Though untethered to direct observations at $z>10$, we can
alternatively estimate gas masses in our sample using the
star-formation surface density to gas mass surface density conversion,
or the Kennicutt-Schmidt (KS) relation
\citep{schmidt59a,kennicutt98a}.  Adopting a unimodal KS relation with
powerlaw index roughly $=2$ \citep{ostriker11a,narayanan12a} and with
measured star-formation surface densities ranging $6<\Sigma_{\rm
  SFR}<100$\,\sfr\,kpc$^{-2}$, we extrapolate that the molecular
(H$_2$) gas masses would range from
$\sim2\times10^{8}-4\times10^{9}$\,\msun.  This presumes
  that the size of gas reservoir is similar to the stellar reservoir.
This would imply molecular gas fractions (H$_2$ to total baryonic
mass) of $\sim40$\%\ on average.  The effect of the added component of
molecular gas on the total baryonic mass volume density is shown in
Figure~\ref{fig:mbk} in open circles.  In the case of the
$z\sim12$ candidates, this shows that summing the stellar mass and
molecular gas mass fully compensates for the predicted baryonic
content of these early halos, leaving little room for other baryonic
contributions, for example, like atomic hydrogen, {\sc HI}, which is
an essential building block of molecular gas and transitional state
for primordial gas to be transformed into stars.

Follow-up ALMA observations of these systems will prove invaluable to
provide an independent estimate of the galaxies' total baryonic
budgets.  For example, spectral scans for \oiii\ (at rest-frame
88\um), will not only provide much-needed spectroscopic confirmation
of these sources but also facilitate a direct dynamical mass estimate,
accounting for both gas and stars.  In the case of efficient star
formation, we might expect the dynamical mass constraint from
\oiii\ to be approximately equal to $f_{\rm b}M_{\rm
  halo}\approx1-2\times10^{10}$\,\msun.  In the case of lower
efficiency and higher halo mass (as one may expect from alternative
cosmological models, see the next section), one would
expect the dynamical mass constraint to be factors of several times
higher due to the overall larger baryonic mass present in the halo.

\subsection{Alternative Cosmologies predict more massive halos early}\label{sec:altcosmology}

An alternative interpretation to the very high stellar baryon fractions
implied in our sample (with $\epsilon_{\star}\approx0.2-0.5$) is that
 the six-parameter $\Lambda$CDM model
  (hereafter $\Lambda$CDM) underestimates the number density of
massive halos at early times.  This revision to the cosmological
framework could be explained through the Early Dark Energy (EDE) model
\citep{karwal16a,poulin18a} which suggests that an early episode of
dark energy injection, near the time of matter-radiation
  equality (followed by $\Lambda$CDM evolution),
could both explain a higher perceived abundance of massive halos at
early times \citep{klypin21a} and resolve recent measurements of the
Hubble tension \citep[e.g.][]{riess22a}.  As discussed in
\citet{boylan-kolchin23a}, the enhanced  matter density,
  power spectrum slope, and $\sigma_{8}$ in EDE could even be used to
explain the high stellar masses measured in some of the most massive
candidates found to-date by {\it JWST} \citep{labbe22a} over much smaller
regions of the sky \citep[though we also note that more recent work
  has suggested downward revision of their stellar mass estimates due
  to the contribution of strong emission lines and/or
  AGN;][]{endsley23a,labbe23a}.  Indeed, EDE predicts the most
profound differences for the most massive halos, which we are
sensitive to detecting in COSMOS-Web; at $z\simgt10$, EDE predicts
$\sim$10$\times$ the number of halos $>10^{11}\,$\msun\ than expected
by $\Lambda$CDM.  Though not directly measurable in the data we
present in this work, future follow-up spectroscopy may be able to
place more meaningful constraints on dynamical masses of these bright
$z>10$ sources, thus give more direct measurements of the abundance of
massive halos at early times.

\subsection{Stochastic Bursts Driving $M_{\rm UV}<-21$ Galaxies in the EoR}

Provided $\Lambda$CDM still holds, the most straightforward
explanation for the presence of such extraordinarily
 massive galaxies at $z>10$ is their rapid growth
through stochastic bursts of star formation where a significant
fraction of available baryons is efficiently cooled, condensed, and
transformed to stars on $<$100\,Myr timescales. The
  similar volume densities measured for very massive galaxies
  ($\sim$10$^{10}$\,\msun) relative to those 10$\times$ less massive
  suggests that very high stellar baryon fractions
  ($\epsilon_\star\sim0.2-0.5$) are not typical of the broader population;
  the steepness of the halo mass function would otherwise demand that
  sources 10$\times$ less massive are $\sim$100$\times$ more
  common. With burst-driven star-formation, galaxies can deviate to
  higher $\epsilon_\star$ and Malmquist bias ensures they are the
  first to be characterized.

This hypothesis is consistent with burst-driven star-formation
dominating the bright end of the UVLF as suggested in cosmological
zoom-in simulations \citep{sun23a}. Such efficient and quick growth
  would be facilitated by the lack of UV background in the
  pre-reionization era.  These bursts may 
    be comprised of several brief ($<$5\,Myr) episodes of
  super-Eddington star formation \citep[e.g.][]{ferrara23a} consistent
  with the Feedback-Free Starbursts model \citep{dekel23a},
    though future dust, gas and spectroscopic observations
    may provide crucial tests for such dust-poor, low-metallicity
    models.  Ideally, direct mass constraints on every luminous
  $z>10$ candidate identified may better inform their star-formation
  histories.
Another key technique that might be used to constrain the
stochasticity of bright galaxies in the EoR is a clustering analysis,
as proposed in \citet{munoz23a} where the $M_{\rm UV}<-21$ population
bias may provide insights for their host halo masses.

If the bright end of the UVLF (e.g. $M_{UV}<-21$) is dominated by
stochastic bursts with intrinsic timescales of
$<$100\,Myr, then a natural consequence
  is an evolution in the shape of the UVLF around
$z\sim8$ from a double power law (at $z>8$) to a Schechter function
(at $z<8$)\footnote{We note that several works have argued the UVLF is
intrinsically a double power law down to $z\sim4$, though such work
primarily relates to \muv\,$\sim$\,$-23$ sources whose luminosity is
likely attributable to AGN
\citep{stevans18a,adams20a,harikane22a,finkelstein22b}, and here we
discuss the $-20\simgt M_{\rm UV}\simgt -22$ regime not thought to be AGN
dominated.}.  This would be the result if a UVLF is calculated in
approximately fixed-width
bins; for example, at $z\simgt8$,  a bin
  of width $\Delta z\approx1$ corresponds to a timescale
less than 100\,Myr, meaning bursts with timescales
  $\sim$50\,Myr will be observed with a duty cycle of order unity
($>$0.5) whereas at $z\sim6$ the duty cycle would be
  substantially lower ($\le$0.25). Conversely, a careful analysis of
the epoch marking the transition between a double powerlaw and
Schechter function may help us directly constrain the characteristic
burst timescale of very bright ($M_{\rm UV}<-21$) galaxies without the
need to invoke dust or feedback to suppress the bright end of the
UVLF.

\section{Conclusions}

We have presented 15 intrinsically luminous candidate galaxies at
$10\simlt z\simlt 14$ with estimated UV absolute magnitudes spanning
$-20.5>M_{\rm UV}>-22$; three are identified as probable low-$z$
contaminants and the remaining 12 are separated in three subsets:
exceptionally bright $10<z<12$ galaxy candidates with $M_{\rm
  UV}<-21.5$, bright $10<z<12$ galaxy candidates spanning
$-20.5>M_{\rm UV}>-21.5$, and $z>13$ candidates with \muv\,$<\,-20.5$
which are only detected in F277W and F444W with more uncertain
physical characteristics.
These sources were identified in the first 0.28\,deg$^2$ area covered
by the COSMOS-Web Survey \citep{casey22a}; their detection is only
made possible by the exquisite sensitivity of the {\it JWST} NIRCam LW
channels.

The rest-frame UV luminosities are among the brightest sources ever
identified at these redshifts, on average 1--3 magnitudes brighter than
other newly identified {\it JWST} $z>10$ galaxy candidates.  Their
rest-frame UV colors are slightly redder as well (with
  $\langle\beta\rangle=-1.7\pm0.5$), perhaps hinting at more complex
underlying star-formation histories or the existence of early dust
reservoirs that redden the stellar continua.  All sources are
spatially resolved with average $R_{\rm eff}\sim500$\,pc. Their
stellar mass surface densities are on par with local elliptical
galaxies and ultra compact dwarf galaxies.  Their star-formation
surface densities are similar to other exceptionally luminous $z>7$
LBG candidates as well as local luminous infrared galaxies.  Their
stellar masses span $10^{8.5}-10^{10}$\,\msun\ with volume densities
roughly of order 10$^{-6}$\,Mpc$^{-3}$.

Four of the 12 robust $z>10$ candidates have UV luminosities similar to GN-z11 at similar or higher
redshifts.  Three of these four sources, \galb, \galc, and \gald, test
the limits of early stellar mass assembly with
$M_{\star}\sim5\times10^{9}$\,\msun\ at $z\sim12$.  Given their
implied stellar mass densities $\sim$10$^{4}$\,\msun\,Mpc$^{-3}$ at
$z\sim12$, we infer that $\sim$20-50\%\ of the baryons in their halos
have been converted into stars ($\epsilon_{\star}\sim0.2-0.5$, where
$\epsilon_{\star}=M_{\star}/(f_{b}M_{\rm halo})$).  This requires either
a very high star-formation efficiency from very early times
($\epsilon_{\rm SF}\sim1$ at $z\simgt16$) with stellar mass growth
that far outpaces dark matter growth of the underlying halos, or
alternatively, a higher abundance of high mass halos that might be
possible in alternatives to $\Lambda$CDM.  We favor the first
explanation, of rapid burst-driven growth in the stellar reservoirs,
making it possible to build $\sim10^{10}$\,\msun\ of stars in less
than 100\,Myr.  Such stochastic episodes of star formation may be
responsible for the underlying shape of the bright end of the UVLF; a
double powerlaw could simply arise at $z>8$ after accounting for the
duty cycle of stochastic bursts at the highest redshifts compared to
$z\sim6-8$.

While we have made a best effort to present secure $z>10$ candidates
in this paper, follow-up spectroscopy is crucial to confirm the
extraordinary nature of these candidates.  The facilities best
equipped for that follow-up work are {\it JWST} itself---which could
give a direct spectrum of the rest-frame UV and optical---and ALMA,
which could be used to constrain the cold gas content in and around
their halos.  These sources could represent the brightest galaxies
{\it JWST} {\it will} find in any field at $z>9$ (unless another large
field-of-view survey like COSMOS-Web is conducted in the future), and
thus they serve as an important laboratory for the formation and
evolution of the first bright galaxies, including the search for
Population III stars, the onset of metal and dust production, as well
as direct constraints on the neutral gas fraction at very early times.

\begin{acknowledgements}
Support for this work was provided by NASA through grant JWST-GO-01727
awarded by the Space Telescope Science Institute, which is operated by
the Association of Universities for Research in Astronomy, Inc., under
NASA contract NAS 5-26555.  CMC, HA, MF, JM, ASL and ORC acknowledges
support from the National Science Foundation through grants
AST-2307006, AST-2009577, and the UT Austin College of Natural
Sciences for support.  CMC also acknowledges support from the Research
Corporation for Science Advancement from a 2019 Cottrell Scholar Award
sponsored by IF/THEN, an initiative of the Lyda Hill
Philanthropies.

The Cosmic Dawn Center (DAWN) is funded by the Danish National
Research Foundation (DNRF) under grant No. 140.
This work was made possible thanks to the CANDIDE cluster at the
Institut d’Astrophysique de Paris, which was funded through grants
from the PNCG, CNES, DIM-ACAV, and the Cosmic Dawn Center; CANDIDE is
maintained by S. Rouberol.
The French contingent of the COSMOS team is partly supported by the
Centre National d'Etudes Spatiales (CNES). OI acknowledges the funding
of the French Agence Nationale de la Recherche for the project iMAGE
(grant ANR-22-CE31-0007).
MBK acknowledges support from NSF CAREER award AST-1752913, NSF grants
AST-1910346 and AST-2108962, NASA grant 80NSSC22K0827, and
HST-AR-15809, HST-GO-15658, HST-GO-15901, HST-GO-15902, HST-AR-16159,
HST-GO-16226, HST-GO-16686, HST-AR-17028, and HST-AR-17043 from the
Space Telescope Science Institute, which is operated by AURA, Inc.,
under NASA contract NAS5-26555.
SG acknowledges financial support from the Villum Young Investigator
grant 37440 and 13160.

Based in part on observations collected at the European Southern
Observatory under ESO programmes 179.A-2005 and 198.A-2003 and on data
obtained from the ESO Science Archive Facility with DOI {\tt
  https://doi.org/10.18727/archive/52}, and on data products produced
by CALET and the Cambridge Astronomy Survey Unit on behalf of the
UltraVISTA consortium.
\end{acknowledgements}

\bibliography{caitlin-bibdesk}

\end{document}

%% file: biofigure.tex
\begin{figure*}
\centering
\includegraphics[width=0.99\columnwidth]{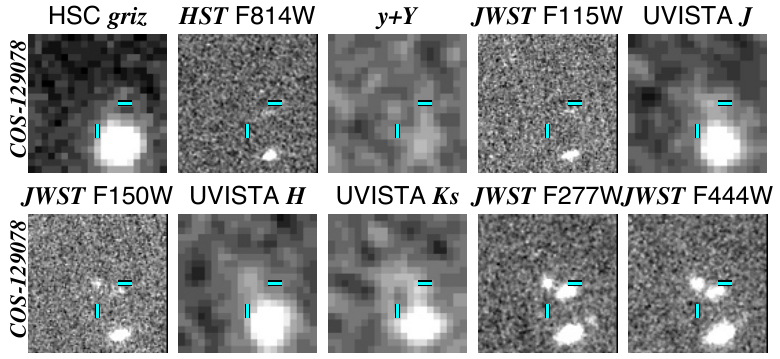}
\includegraphics[width=0.99\columnwidth]{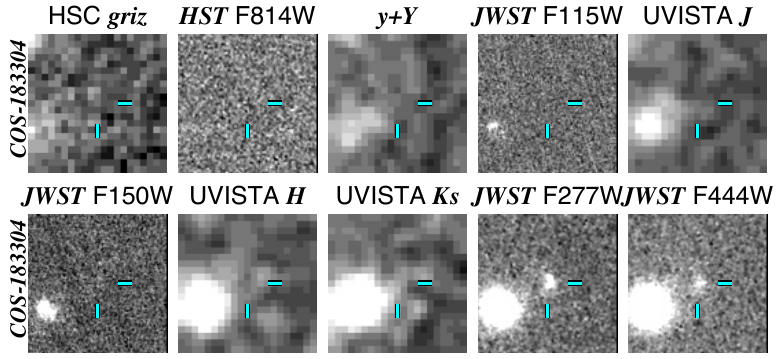}\\
\includegraphics[width=0.99\columnwidth]{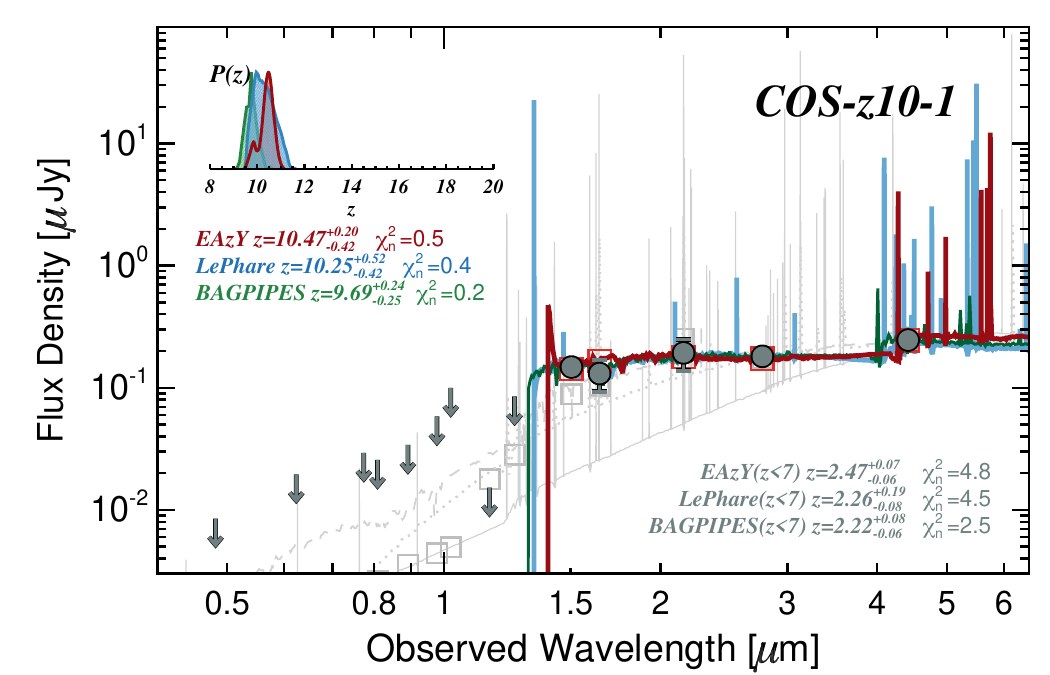}
\includegraphics[width=0.99\columnwidth]{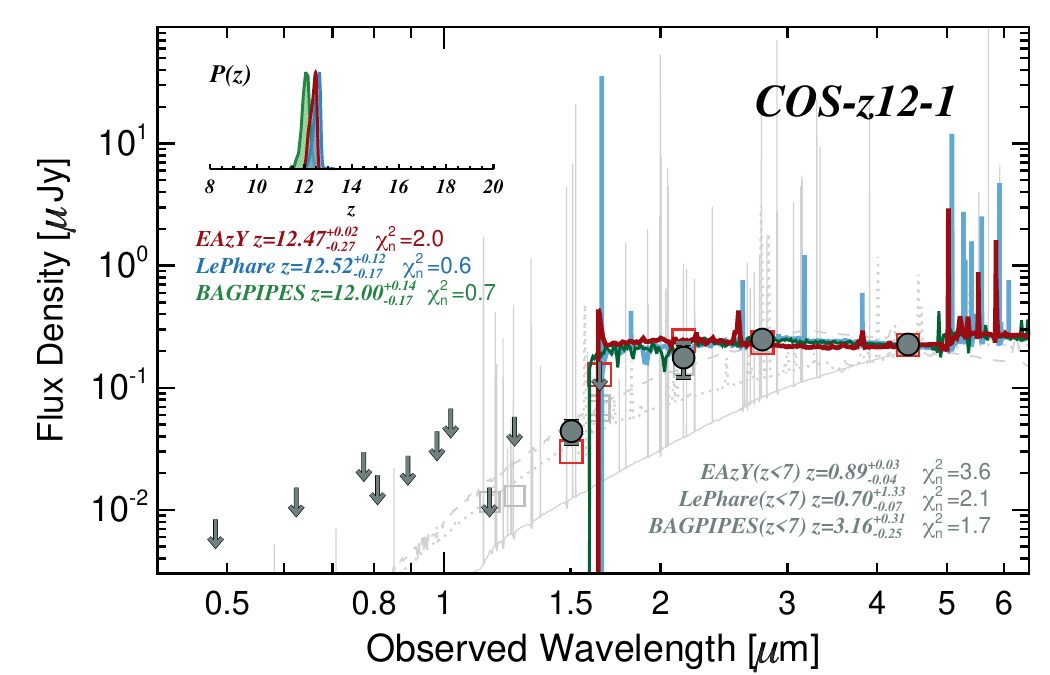}\\
\includegraphics[width=0.99\columnwidth]{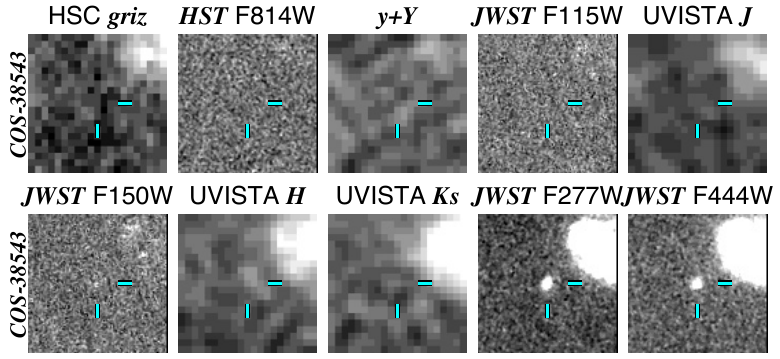}
\includegraphics[width=0.99\columnwidth]{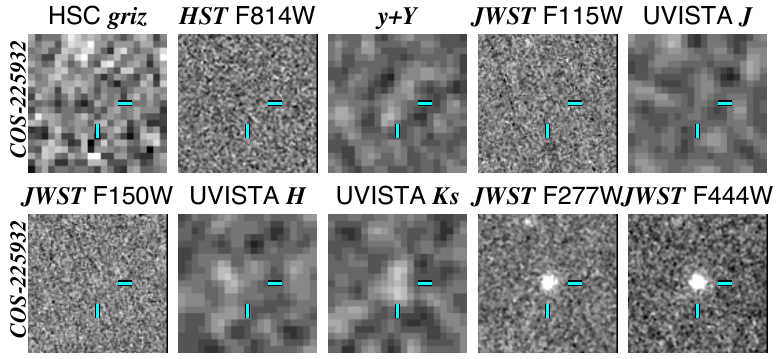}\\
\includegraphics[width=0.99\columnwidth]{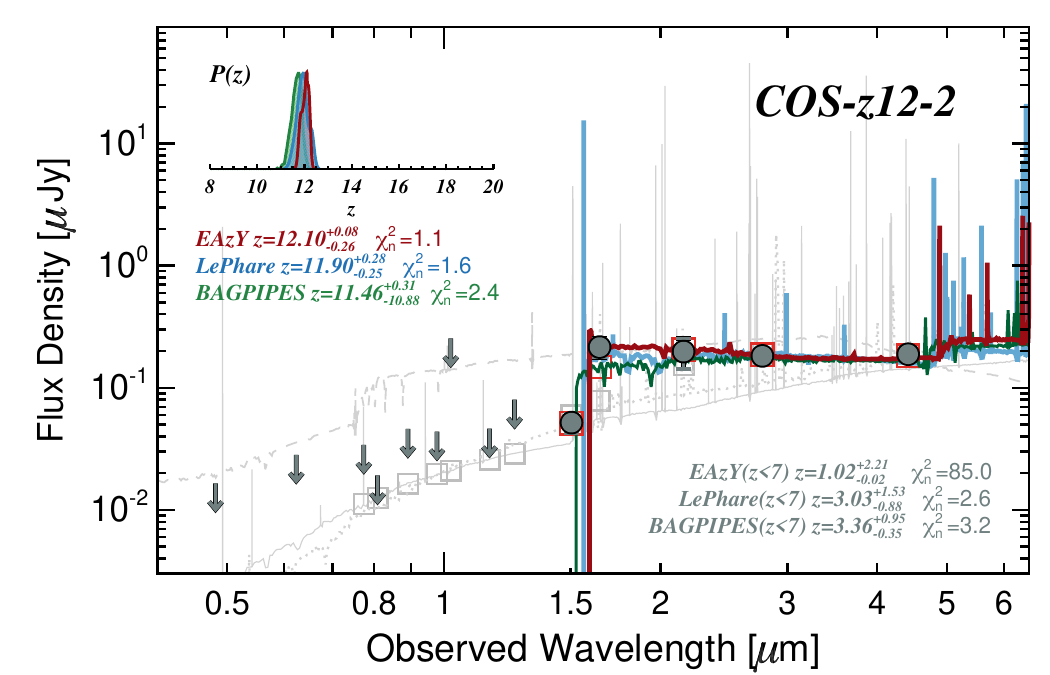}
\includegraphics[width=0.99\columnwidth]{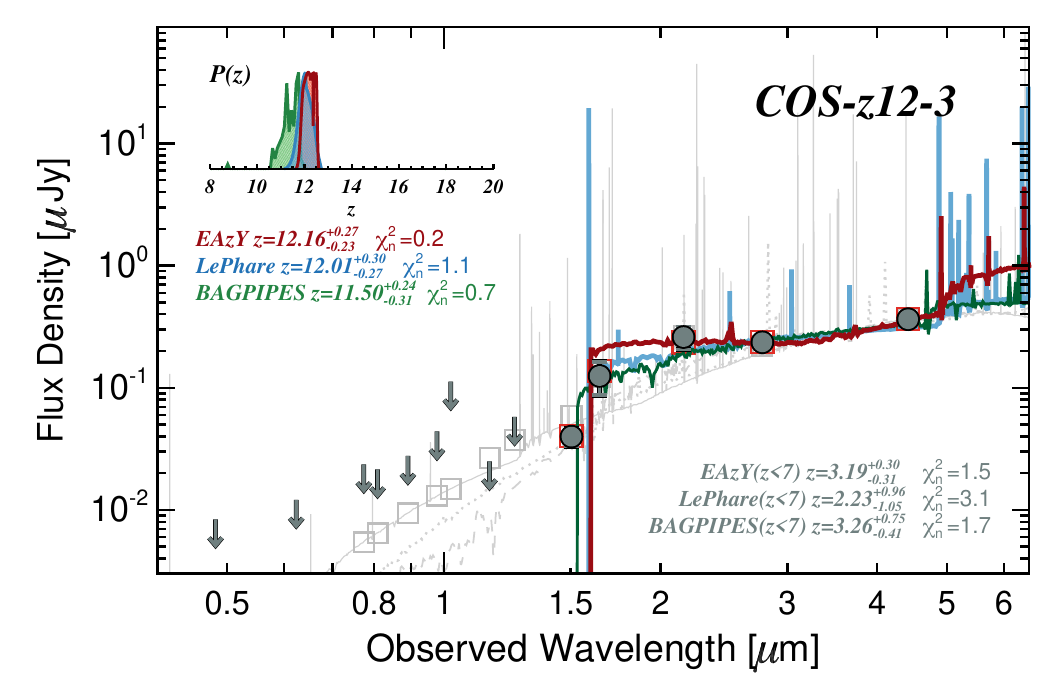}\\
\caption{Cutouts and SEDs of all sources in our sample. Cutouts are
  $3''\times3''$ and include, from left to right: a stack of HSC
  $griz$, {\it Hubble (HST)} F814W, a stack of UltraVISTA $Y$ and HSC $y$,
  {\it JWST} F115W, UltraVISTA $J$, {\it JWST} F150W, UltraVISTA $H$ and $Ks$, and
  {\it JWST} F277W and F444W.  The plotted SED shows model-based photometry
  and 2$\sigma$ upper limits for photometric points below 3$\sigma$
  significance (gray).  We overplot best-fit high-redshift solutions
  from EAzY (red), LePhare (blue) and BAGPIPES (green), and their
  corresponding redshift probability density distributions between
  $8<z<20$ on the inset plot.  In gray we show the best-fit redshift
  solutions forced to $z<7$ from EAzY (dashed), LePhare (solid), and
  BAGPIPES (dotted).  Synthesized photometry are shown in open boxes
  for two SEDs only for clarity: the EAzY-based high-$z$ solution
  (red) and the LePhare low-$z$ solution (gray). Normalized
  $\chi^2_{n}$ values {\color{blue}(see \S~\ref{sec:chi2})} are given
  in each panel for both high-$z$ and low-$z$ solutions.}
\label{fig:bio1}
\end{figure*}
\onecolumngrid
\clearpage
  \includegraphics[width=0.49\columnwidth]{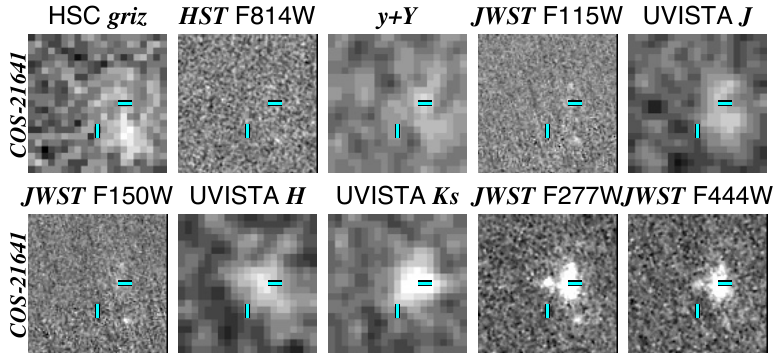}
\includegraphics[width=0.49\columnwidth]{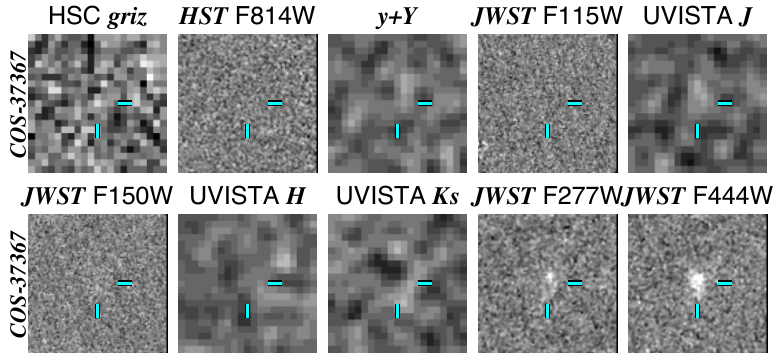}
\includegraphics[width=0.49\columnwidth]{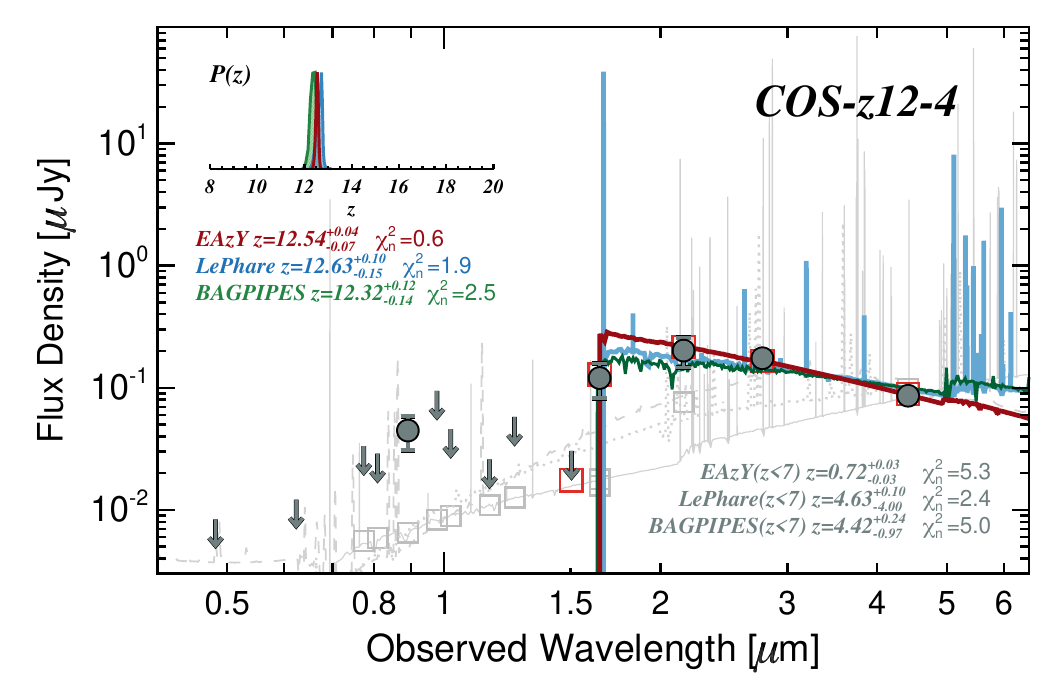}
\includegraphics[width=0.49\columnwidth]{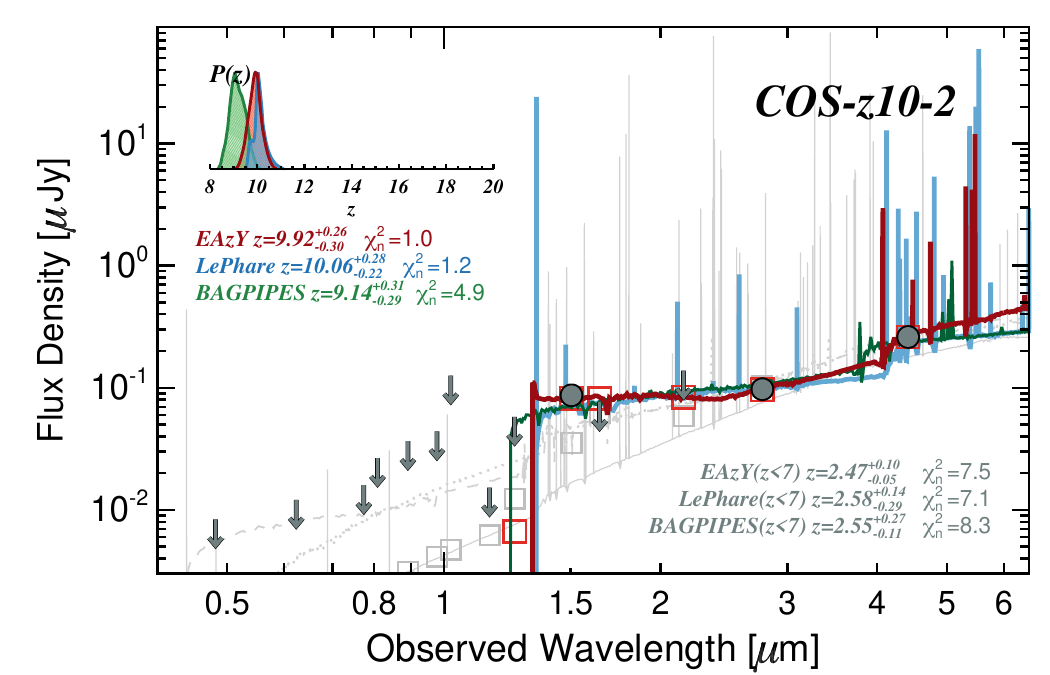}
  \includegraphics[width=0.49\columnwidth]{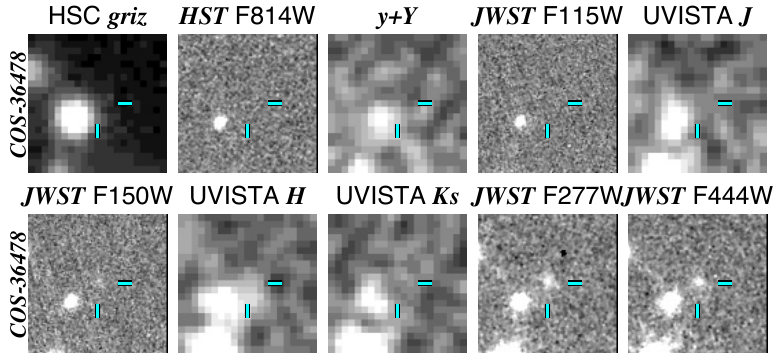}
\includegraphics[width=0.49\columnwidth]{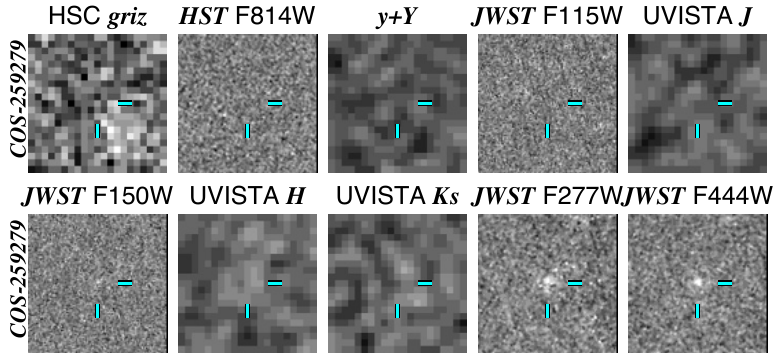}
\includegraphics[width=0.49\columnwidth]{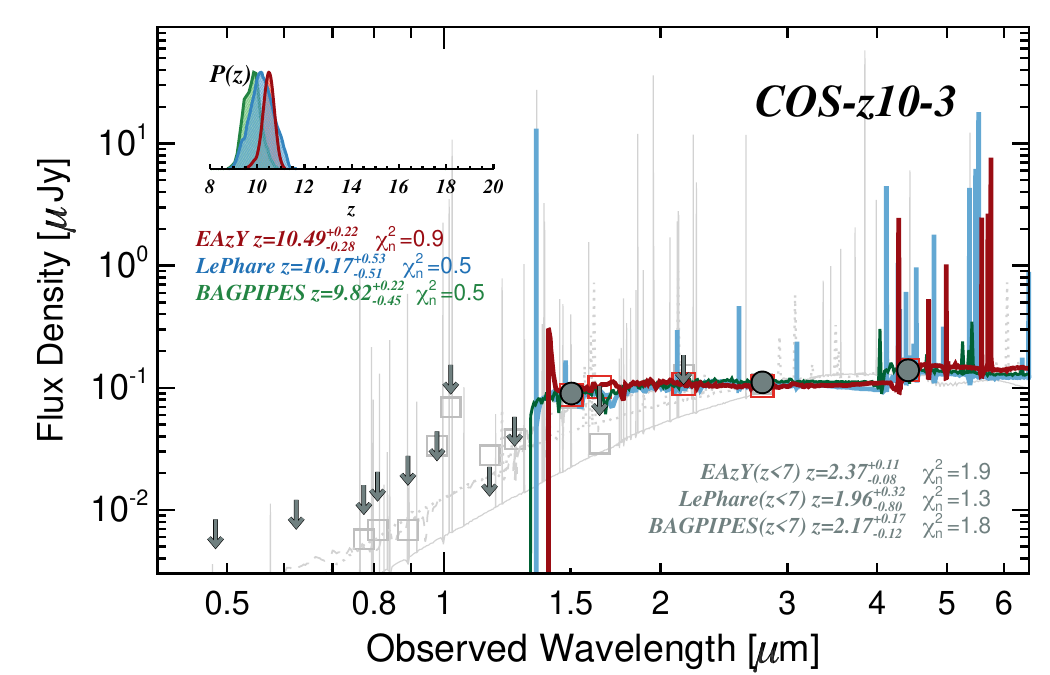}
\includegraphics[width=0.49\columnwidth]{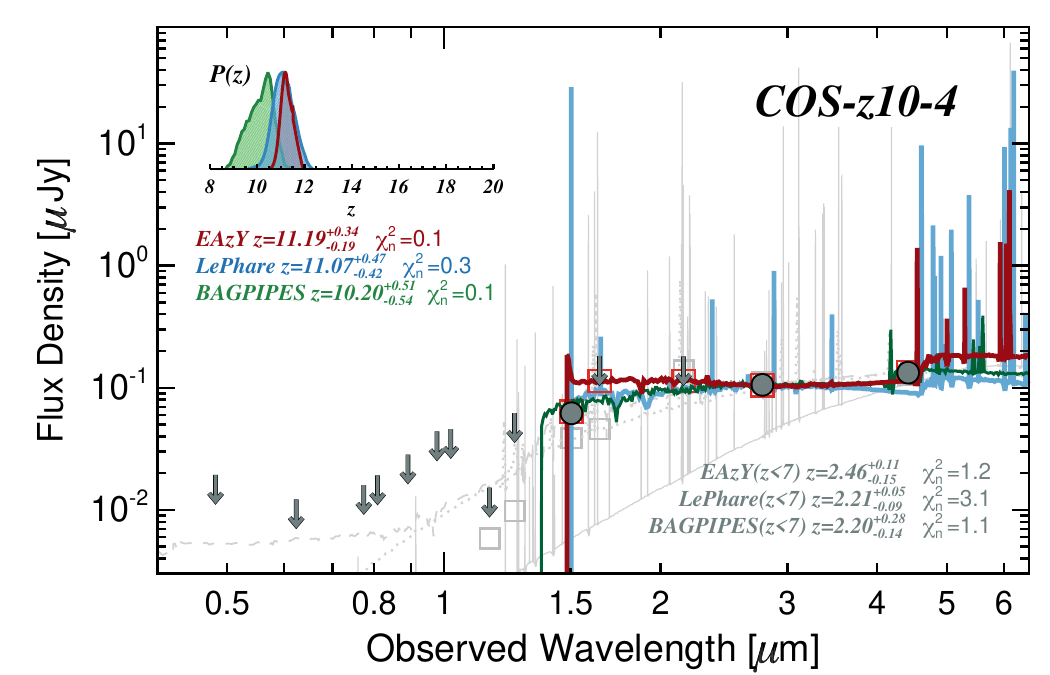}
\centerline{{\small Figure~\ref{fig:bio1} --- continued.}}
\clearpage
\includegraphics[width=0.49\columnwidth]{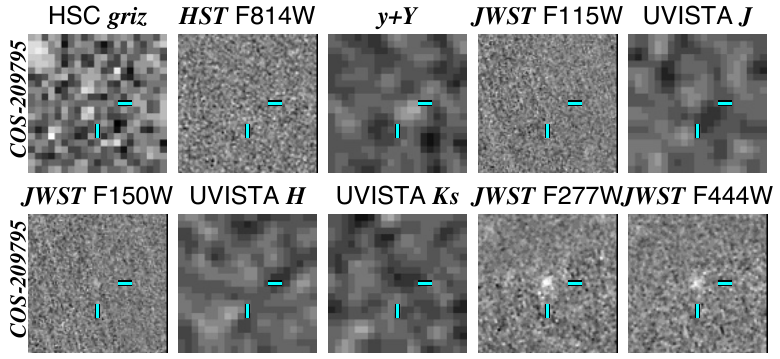}
\includegraphics[width=0.49\columnwidth]{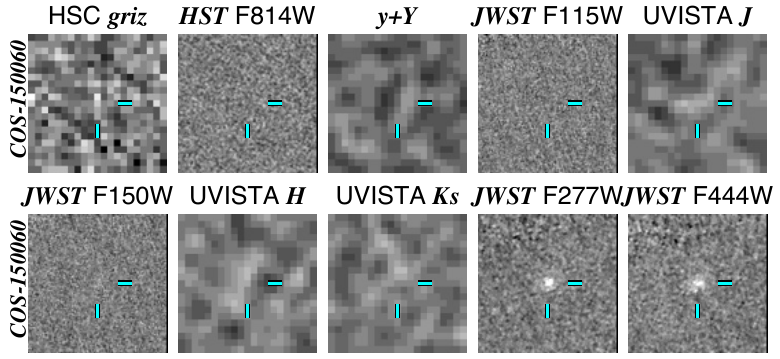}
\includegraphics[width=0.49\columnwidth]{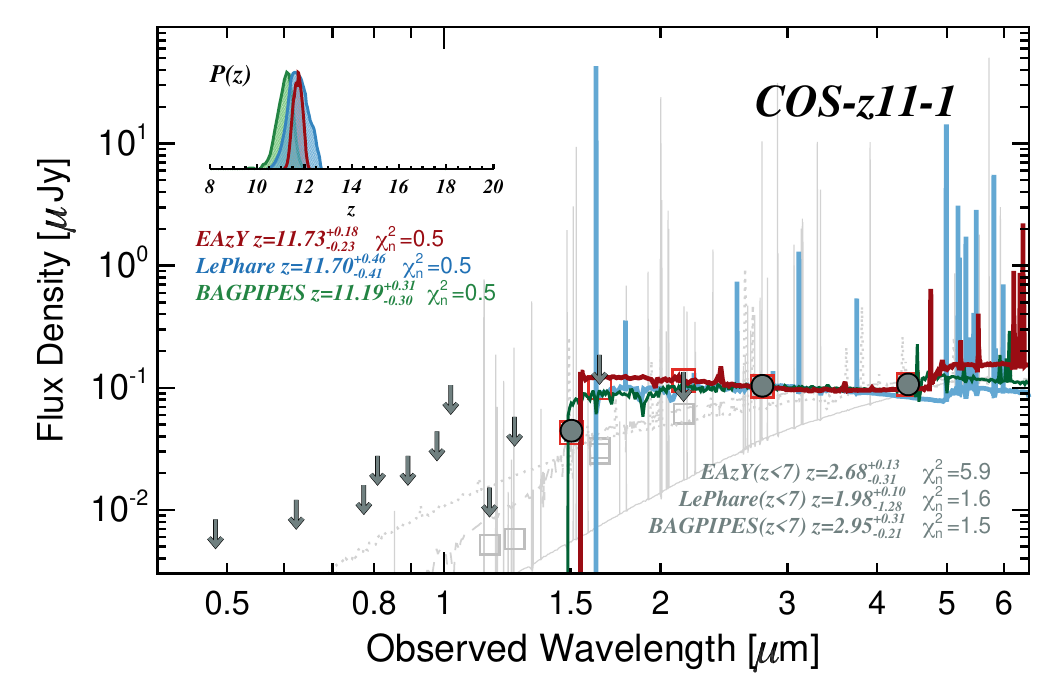}
\includegraphics[width=0.49\columnwidth]{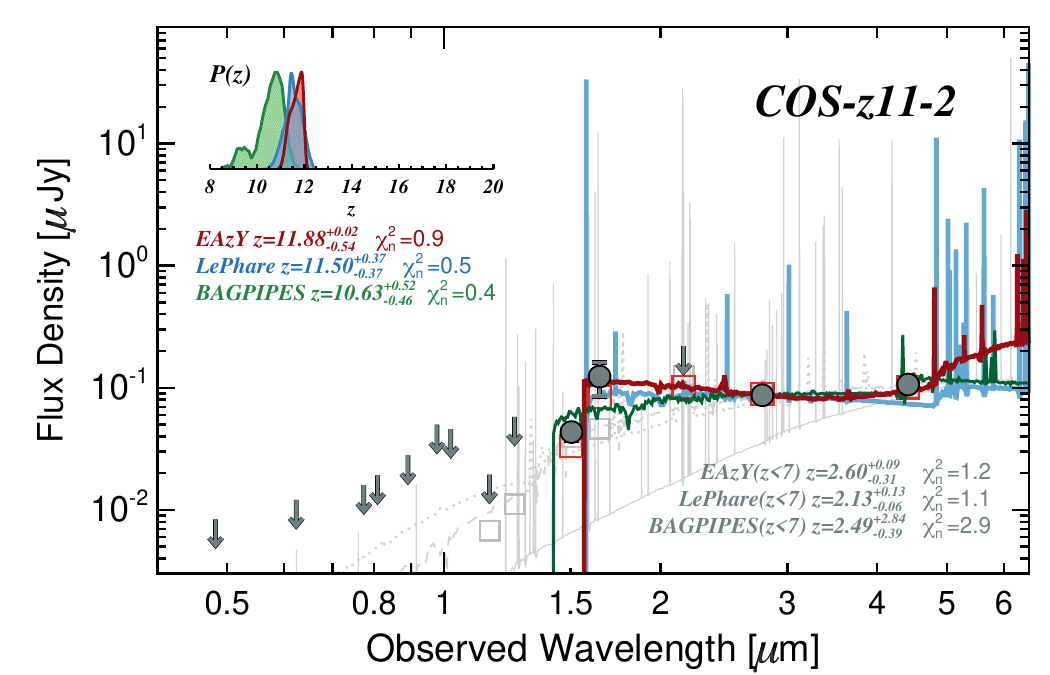}
  \includegraphics[width=0.49\columnwidth]{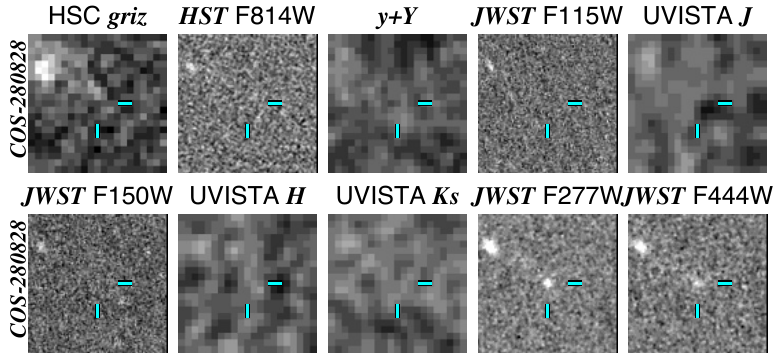}
\includegraphics[width=0.49\columnwidth]{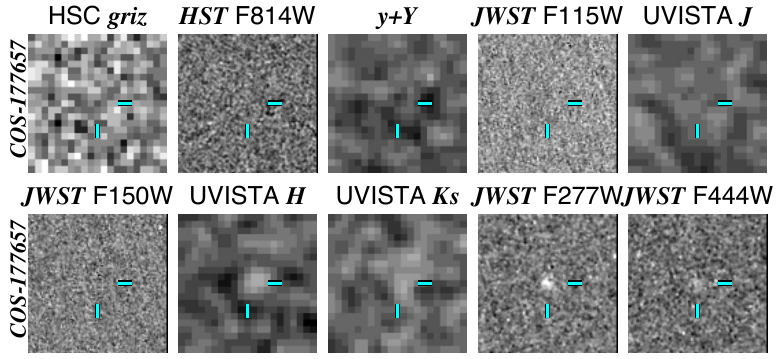}
\includegraphics[width=0.49\columnwidth]{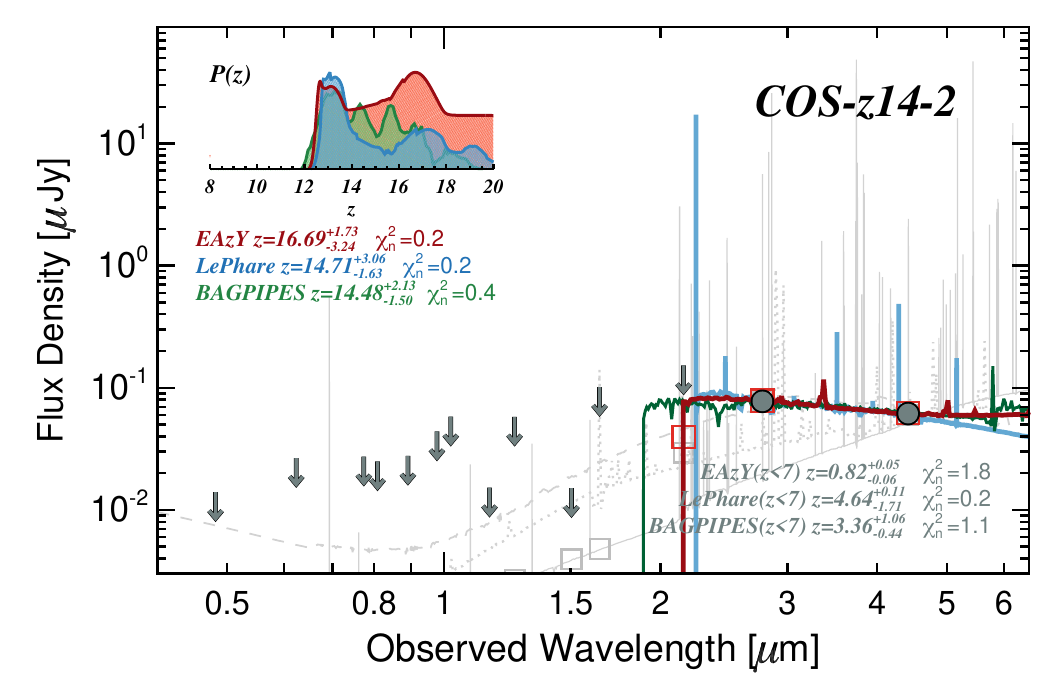}
\includegraphics[width=0.49\columnwidth]{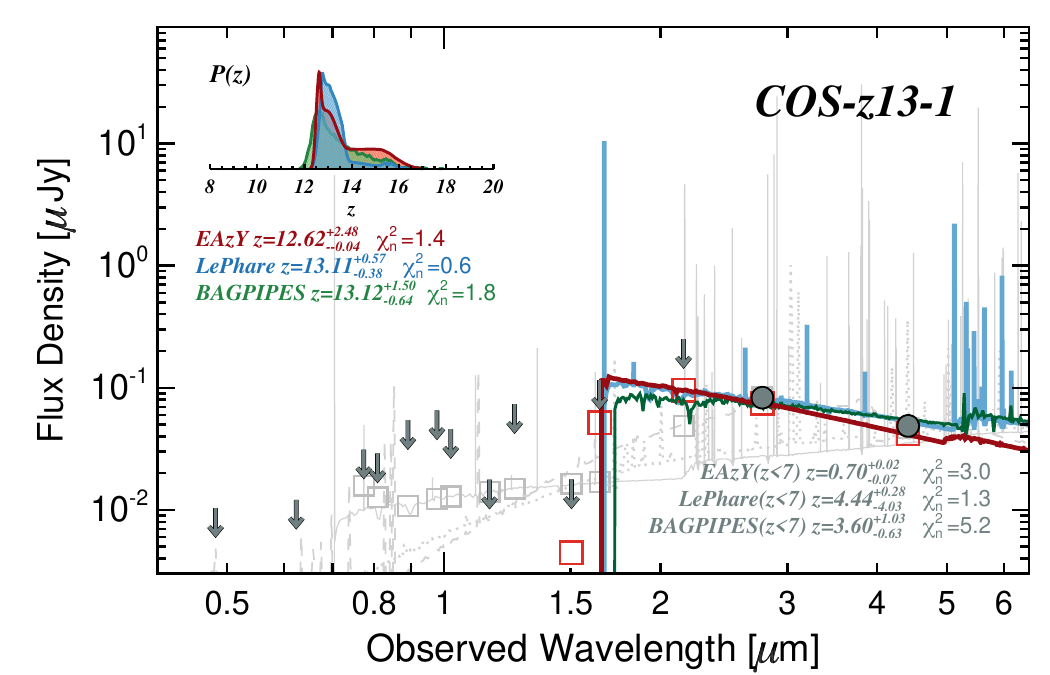}
\centerline{{\small Figure~\ref{fig:bio1} --- continued.}}
\clearpage
{\centering
\includegraphics[width=0.49\columnwidth]{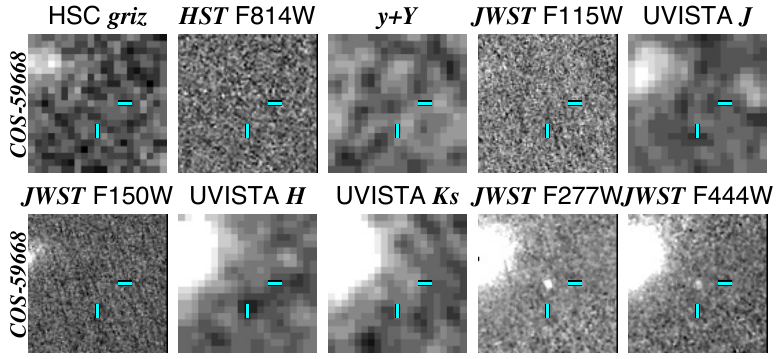}
\includegraphics[width=0.49\columnwidth]{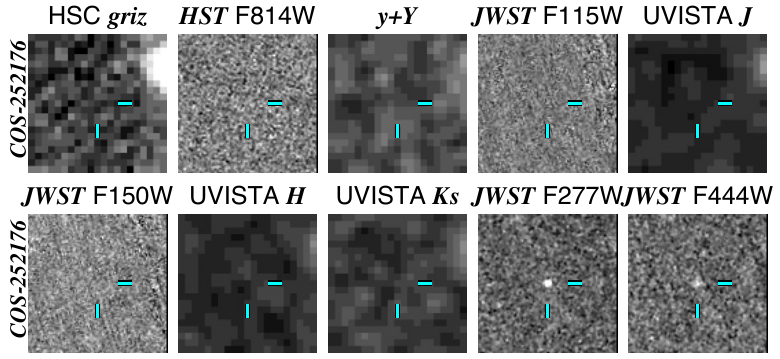}
\includegraphics[width=0.49\columnwidth]{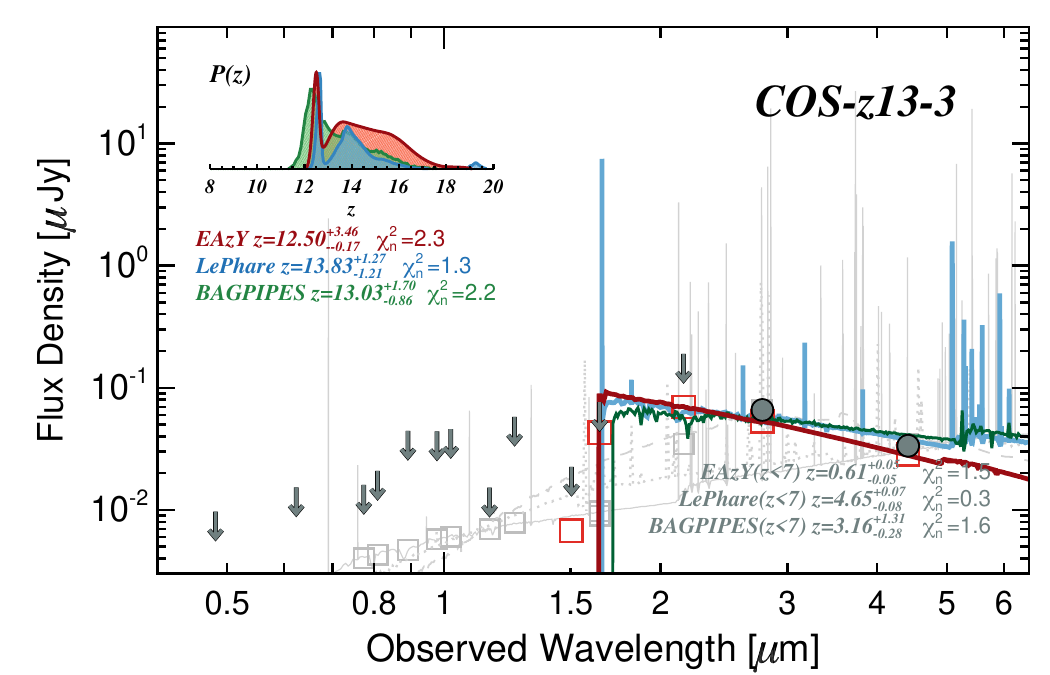}
\includegraphics[width=0.49\columnwidth]{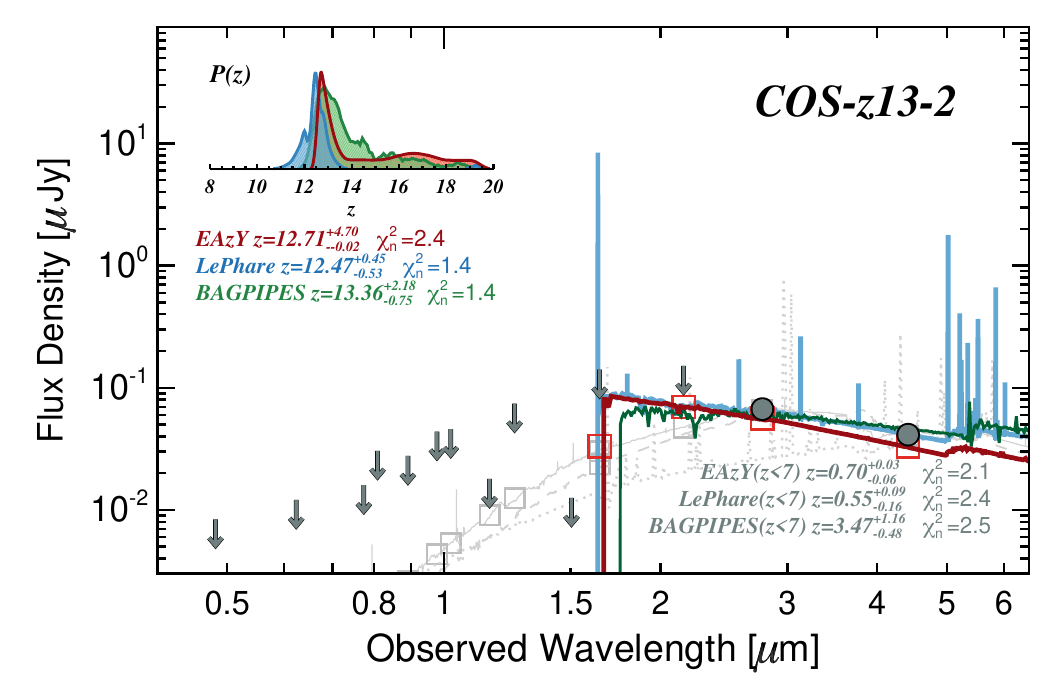}
\includegraphics[width=0.49\columnwidth]{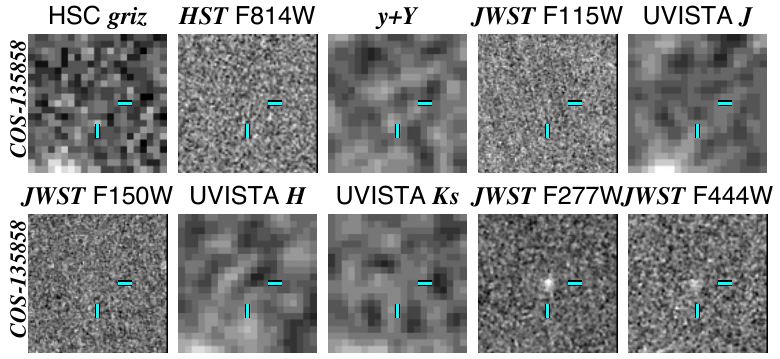}\\
\includegraphics[width=0.49\columnwidth]{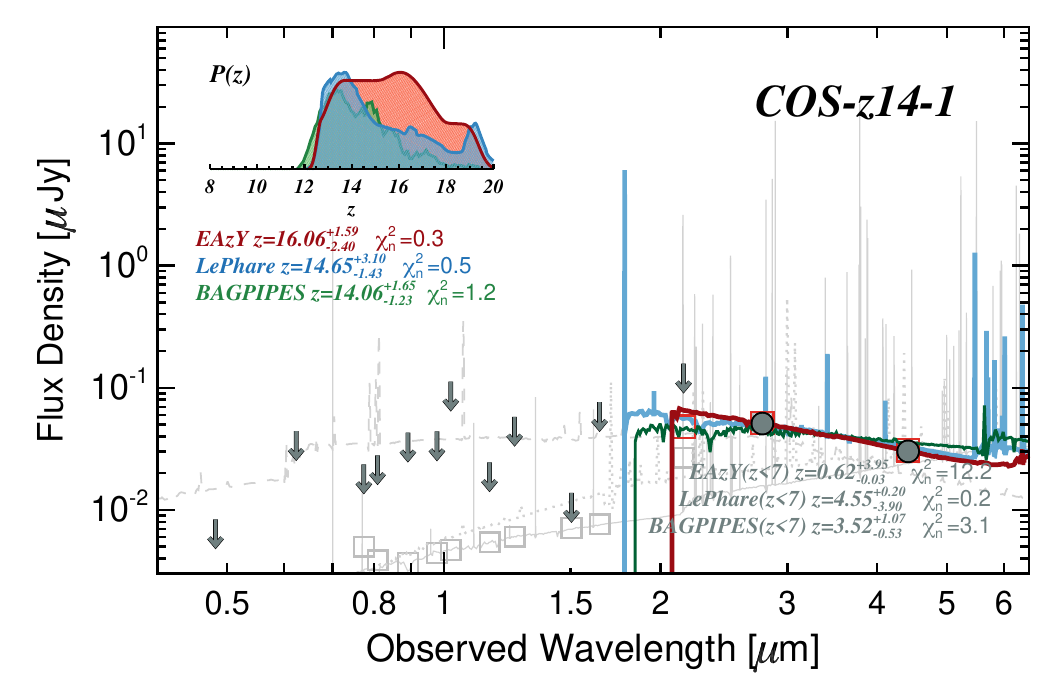}\\
\centerline{{\small Figure~\ref{fig:bio1} --- continued.}}}
\vspace{5mm}
\clearpage
\twocolumngrid

%% file: tab_phot.tex
\begin{longrotatetable}
\begin{deluxetable*}{@{\extracolsep{4pt}}c@{}c@{}c@{}c@{}c@{}c@{}c@{}c@{}c@{}c@{}c@{}c@{}c@{}c@{}c}
  \tablecaption{$10<z<14$ Galaxy Candidate Photometry\label{tab:phot}}
  \tablewidth{670pt}
  \tabletypesize{\scriptsize}
  \tablehead
      {
    \colhead{}&
    \multicolumn{5}{c}{{\tt SE} {\sc Classic Aperture-Based Photometry}} & 
    \multicolumn{9}{c}{{\tt SE++} {\sc Model-Based Photometry}} \\
    \cline{2-6} \cline{7-15} 
\colhead{{\sc Source}}&    \colhead{F814W} & \colhead{F115W} & \colhead{F150W} & \colhead{F277W} & \colhead{F444W} &
    \colhead{F814W} & \colhead{F115W} & \colhead{F150W} & \colhead{F277W} & \colhead{F444W} &
    \colhead{UVISTA $Y$} & \colhead{UVISTA $J$} & \colhead{UVISTA $H$} & \colhead{UVISTA $Ks$} \\
    \colhead{} & [nJy] & [nJy] & [nJy] & [nJy] & [nJy] &
            [nJy] & [nJy] & [nJy] & [nJy] &  [nJy] & [nJy] & [nJy] & [nJy] & [nJy] \\
  }
\startdata
\gala & 0.5$\pm$9.6  &  $-$8.4$\pm$8.6  &  60.1$\pm$7.1  &  67.4$\pm$3.5  &   92.3$\pm$3.9 & 6.5$\pm$13.4  &    0.0$\pm$9.0  &    147.8$\pm$14.3  &    180.9$\pm$6.9  &    246.0$\pm$8.1  &   0.1$\pm$98.4  &    27.3$\pm$45.3  &    130.7$\pm$61.6  &    192.9$\pm$84.2 \\
\galb & $-$1.7$\pm$9.6  &  $-$4.9$\pm$7.6  &  12.1$\pm$6.3  &  56.2$\pm$3.5  &   47.3$\pm$3.9 & 0.0$\pm$10.1  &    0.0$\pm$8.6  &    44.1$\pm$10.9  &    248.2$\pm$6.0  &    226.2$\pm$7.1  &    21.6$\pm$30.9  &    0.0$\pm$33.5  &    87.4$\pm$49.7  &    176.7$\pm$68.3 \\
\galc & $-$0.1$\pm$9.6  &  1.2$\pm$7.8  &  26.2$\pm$6.3  &  82.0$\pm$3.5  &   80.3$\pm$3.9 & 0.0$\pm$9.6  &    21.9$\pm$14.4  &    51.8$\pm$11.0  &    182.4$\pm$6.0  &    188.2$\pm$7.3  & 124.3$\pm$106.8  &    22.0$\pm$45.9  &    215.4$\pm$62.2  &    197.7$\pm$85.5 \\
\gald & 0.5$\pm$9.6  &  $-$0.3$\pm$7.9  &  18.7$\pm$6.4  &  94.2$\pm$3.5  &   142.5$\pm$3.9 &  2.3$\pm$11.8  &    8.4$\pm$11.2  &    39.8$\pm$9.2  &    235.8$\pm$6.3  &    364.2$\pm$8.5  &   39.0$\pm$92.5  &    0.0$\pm$39.2  &    124.9$\pm$55.5  &    260.3$\pm$76.1 \\
\galn &  0.5$\pm$9.6  &  $-$2.0$\pm$8.2  &  22.5$\pm$6.7  &  29.3$\pm$3.5  &   89.7$\pm$3.9 &  7.5$\pm$11.1  &    0.0$\pm$10.3  &    86.8$\pm$11.5  &    97.4$\pm$5.9  &    260.5$\pm$7.7  &    65.2$\pm$90.2  &    0.0$\pm$999.8  &    0.0$\pm$1000.4  &    27.9$\pm$75.7 \\
\gale & 0.4$\pm$9.6  &  $-$0.3$\pm$7.7  &  37.1$\pm$6.5  &  46.3$\pm$3.5  &   58.4$\pm$3.9 & 1.4$\pm$12.1  &    0.3$\pm$15.1  &    89.7$\pm$12.5  &    110.7$\pm$6.9  &    139.1$\pm$8.6  &    67.8$\pm$95.2  &    0.0$\pm$39.2  &    26.8$\pm$57.3  &    75.0$\pm$78.6 \\
\galf & $-$0.5$\pm$9.6  &  $-$4.5$\pm$8.0  &  24.3$\pm$6.3  &  41.0$\pm$3.5  &   47.1$\pm$3.9  &  0.0$\pm$9.6  &    0.0$\pm$10.3  &    61.8$\pm$10.2  &    105.6$\pm$6.0  &    132.9$\pm$6.9  &    0.0$\pm$28.4  &    4.1$\pm$34.7  &    105.1$\pm$48.6  &    70.8$\pm$66.8 \\
\galh & 1.0$\pm$9.6  &  9.2$\pm$8.5  &  20.6$\pm$6.8  &  43.5$\pm$3.5  &   44.5$\pm$3.9 & 8.5$\pm$11.1  &    0.0$\pm$10.8  &    44.5$\pm$10.9  &    103.7$\pm$5.8  &    106.6$\pm$6.8  & 36.1$\pm$91.7 &  0.0$\pm$39.3  &    109.4$\pm$55.4  &    24.2$\pm$75.8 \\
\galg &  $-$0.4$\pm$9.6  &  $-$2.4$\pm$7.6  &  20.8$\pm$6.3  &  39.2$\pm$3.5  &   45.7$\pm$3.9  &  0.0$\pm$10.1  &    4.3$\pm$10.0  &    43.4$\pm$8.2  &    86.8$\pm$4.4  &    106.4$\pm$5.1  & 0.0$\pm$84.9  &    0.0$\pm$39.3  &    123.5$\pm$55.1  &    109.5$\pm$75.6 \\
\galj & 0.3$\pm$9.6  &  $-$4.7$\pm$8.0  &  0.0$\pm$7.3  &  44.6$\pm$3.5  &   27.1$\pm$3.9 & 9.8$\pm$10.4  &    2.5$\pm$11.8  &    5.3$\pm$9.8  &    82.9$\pm$5.9  &    48.7$\pm$6.3  &     0.0$\pm$28.9  &    15.6$\pm$34.4  &    39.4$\pm$48.2  &    140.4$\pm$66.0 \\
\gall & 1.5$\pm$9.6  &  $-$8.0$\pm$9.7  &  2.4$\pm$7.5  &  44.9$\pm$3.5  &   32.2$\pm$3.9 & 11.3$\pm$10.0  &    2.7$\pm$10.8  &    0.0$\pm$8.3  &    66.9$\pm$5.0  &    41.3$\pm$5.8  &  0.0$\pm$85.2  &    16.4$\pm$40.8  &    64.8$\pm$55.1  &    41.0$\pm$75.4 \\
\galm & 0.7$\pm$9.6  &  2.9$\pm$7.6  &  3.4$\pm$6.3  &  27.8$\pm$3.5  &   21.1$\pm$3.9 &  8.9$\pm$10.7  &    9.3$\pm$8.9  &    1.2$\pm$7.3  &    50.9$\pm$4.5  &    30.2$\pm$4.8  & 46.0$\pm$91.1  &    0.0$\pm$39.2  &    0.1$\pm$55.5  &    47.5$\pm$76.7 \\
\galz & 0.5$\pm$9.6  &  10.9$\pm$7.6  &  10.5$\pm$6.3  &  80.6$\pm$3.5  &   45.2$\pm$3.9 & 9.7$\pm$12.0  &    10.7$\pm$11.5  &    17.9$\pm$9.4  &    174.6$\pm$12.8  &    86.0$\pm$8.8  &    0.0$\pm$87.8  &    0.0$\pm$39.3  &    120.2$\pm$56.3  &    201.7$\pm$77.2 \\
\galk & 0.0$\pm$9.6  &  $-$4.2$\pm$7.6  &  5.2$\pm$6.3  &  41.4$\pm$3.5  &   27.6$\pm$3.9 &  1.7$\pm$10.1  &    0.0$\pm$10.2  &    10.0$\pm$8.7  &    65.8$\pm$5.3  &    33.5$\pm$5.7  &   0.0$\pm$85.7  &    0.0$\pm$39.3  &    0.0$\pm$53.8  &    79.0$\pm$75.6 \\
\gali & 0.6$\pm$9.6  &  $-$1.5$\pm$7.6  &  8.4$\pm$6.3  &  34.2$\pm$3.5  &   27.6$\pm$3.9 &  5.8$\pm$10.6  &    0.0$\pm$7.6  &    2.5$\pm$7.1  &    77.4$\pm$4.7  &    61.5$\pm$5.1  &     12.0$\pm$29.9  &    0.1$\pm$34.8  &    25.1$\pm$48.8  &    42.1$\pm$67.0 \\
\hline
\multicolumn{15}{c}{{\sc Effective Wavelength of Filters} [\um]}\\
\hline
& 0.814 & 1.15 & 1.50 & 2.77 & 4.44 & 0.814 & 1.15 & 1.50 & 2.77 & 4.44 &  1.02 & 1.25 & 1.65 & 2.15 \\
\enddata
\tablecomments{ The photometry for all of our $z\simgt10$ candidates
  extracted using two different methods: {\tt SE} classic, which is
  aperture-based photometry measured in 0$\farcs$30 diameter apertures
  on PSF-homogenized imaging for space-based data only, and from {\tt
    SE++} single Sersic model-based photometry measured on images in their native resolution.}.  
\end{deluxetable*}
\end{longrotatetable}

%% file: tab_phys.tex
\begin{deluxetable*}{cc@{ }cc@{ }c@{ }c@{ }c@{ }c@{ }c@{ }c@{ }c@{ }c}
  \tabletypesize{\footnotesize}
  \tablecolumns{11}
  \tablecaption{Sample Characteristics}
  \tablehead{
    \colhead{Source} & \colhead{$\alpha_{\rm J2000}$} & \colhead{$\delta_{\rm J2000}$} & \colhead{$z_{\rm phot}$} & \colhead{$z_{\rm phot}$} & \colhead{$z_{\rm phot}$} &
    \colhead{$M_{\rm UV}$} & $\beta_{\rm UV}$ & SFR$_{\rm 100Myr}$ & $M_\star$ &  R$_{\rm eff}$({\sc galfit}) \\
     & & & {\sc Bagpipes} & {\sc EAzY} & {\sc LePhare} & & & [$M_\odot$\,yr$^{-1}$] & [M$_\odot$] & [pc]\\
%
    \vspace{-3mm}
  }
\startdata
\multicolumn{11}{c}{{\sc Super Bright $10<z<12$ Sample}} \\
\gala\ & 10:01:26.00 & $+$01:55:59.70 & 9.69$^{+0.24}_{-0.25}$  & 10.47$^{+0.20}_{-0.42}$ & 10.27$^{+0.52}_{-0.42}$ & $-$21.53$^{+0.10}_{-0.10}$ & $-$1.67$^{+0.14}_{-0.21}$ & 30$^{+8}_{-8}$ & (3.7$^{+2.2}_{-1.4}$)$\times10^{9}$ &  520$\pm$50 \\
\galb\ & 09:58:55.21 & $+$02:07:16.77 & 12.08$^{+0.13}_{-0.16}$ & 12.47$^{+0.02}_{-0.27}$ & 12.54$^{+0.12}_{-0.17}$ & $-$22.19$^{+0.10}_{-0.17}$ & $-$1.78$^{+0.23}_{-0.24}$ & 39$^{+10}_{-16}$ & (4.0$^{+1.4}_{-1.3}$)$\times10^{9}$ & 420$\pm$70 \\
\galc\ & 09:59:59.91 & $+$02:06:59.90 & 11.46$^{+0.43}_{-0.04}$ & 12.10$^{+0.08}_{-0.26}$ & 11.92$^{+0.28}_{-0.26}$ & $-$21.89$^{+0.14}_{-0.15}$ & $-$1.86$^{+0.15}_{-0.27}$ & 31$^{+22}_{-12}$ & (4.8$^{+5.3}_{-2.3}$)$\times10^{9}$ & 450$\pm$30 \\
\gald\ & 09:59:49.04 & $+$01:53:26.19 & 11.46$^{+0.28}_{-0.23}$ & 12.16$^{+0.27}_{-0.23}$ & 12.03$^{+0.30}_{-0.27}$ & $-$21.58$^{+0.07}_{-0.31}$ & $-$0.60$^{+0.22}_{-0.33}$ & 54$^{+16}_{-16}$ & (5.6$^{+1.7}_{-1.4}$)$\times10^{9}$ & 520$\pm$40 \\
\multicolumn{11}{c}{{\sc Bright $10<z<12$ Sample}} \\
\galn & 09:59:51.77 & $+$02:07:15.02 & 9.15$^{+0.29}_{-0.35}$ & 9.92$^{+0.26}_{-0.30}$ & 10.06$^{+0.28}_{-0.22}$ & $-$20.62$^{+0.14}_{-0.17}$ & $-$1.22$^{+0.22}_{-0.16}$ & 44$^{+13}_{-20}$ & (1.1$^{+0.7}_{-0.3}$)$\times10^{10}$ & 1120$\pm$330 \\
\gale\ & 09:59:57.50 & $+$02:06:20.06 & 9.82$^{+0.22}_{-0.45}$ & 10.49$^{+0.22}_{-0.28}$ & 10.17$^{+0.53}_{-0.51}$ & $-$20.97$^{+0.17}_{-0.12}$ & $-$1.61$^{+0.18}_{-0.17}$ & 18$^{+7}_{-4}$ & (1.8$^{+0.7}_{-0.4}$)$\times10^{9}$ & 890$\pm$90 \\
\galf\ & 10:00:37.96 & $+$01:49:32.43 & 10.20$^{+0.51}_{-0.54}$ & 11.19$^{+0.34}_{-0.19}$ & 11.07$^{+0.47}_{-0.42}$ & $-$20.85$^{+0.27}_{-0.19}$ & $-$1.40$^{+0.26}_{-0.23}$ & 18$^{+7}_{-6}$ & (1.9$^{+1.0}_{-0.4}$)$\times10^{9}$ & 870$\pm$90 \\
\galh\ & 09:59:52.53 & $+$02:00:23.53 & $11.19^{+0.31}_{-0.30}$ & 11.73$^{+0.18}_{-0.23}$ & 11.70$^{+0.46}_{-0.41}$ & $-$21.13$^{+0.15}_{-0.21}$ & $-$1.74$^{+0.32}_{-0.17}$ & 16$^{+5}_{-6}$ & (1.7$^{+0.9}_{-0.6}$)$\times10^{9}$ & 620$\pm$70 \\
\galg\ & 10:01:34.80 & $+$02:05:41.48 & 10.63$^{+0.52}_{-0.46}$ & 11.88$^{+0.02}_{-0.54}$ & 11.50$^{+0.37}_{-0.37}$ & $-$20.77$^{+0.23}_{-0.34}$ & $-$1.48$^{+0.27}_{-0.39}$ & 15$^{+4}_{-9}$ & (1.6$^{+0.9}_{-0.5}$)$\times10^{9}$ & 520$\pm$90 \\
\multicolumn{11}{c}{{\sc Sample at $z>13$}} \\
\galj\ & 09:59:05.75 & $+$02:04:04.39 & 13.2$^{+0.6}_{-0.9}$ & 12.6$^{+2.5}_{-0.1}$ & 13.1$^{+0.6}_{-0.4}$ & $-$21.27$^{+0.16}_{-0.14}$ & $-$2.44$^{+0.15}_{-0.19}$ & 5.6$^{+2.7}_{-3.3}$ & (5.9$^{+5.0}_{-2.1}$)$\times10^{8}$ & 450$\pm$50 \\
\gall\ & 10:00:04.24 & $+$02:02:11.19 & 13.4$^{+0.7}_{-1.2}$ & 12.7$^{+4.7}_{-0.1}$ & 12.5$^{+0.5}_{-0.5}$ & $-$21.03$^{+0.17}_{-0.17}$ & $-$2.37$^{+0.19}_{-0.14}$ & 5.4$^{+1.9}_{-3.1}$ & (5.6$^{+3.4}_{-2.2}$)$\times10^{9}$ & 190$\pm$30 \\
\galm\ & 10:01:31.17 & $+$01:58:45.00 & 14.0$^{+1.1}_{-1.5}$ & 16.1$^{+1.6}_{-2.4}$ & 14.7$^{+3.1}_{-1.4}$ & $-$20.75$^{+0.15}_{-0.19}$ & $-$2.35$^{+0.19}_{-0.16}$ & 5.0$^{+1.4}_{-2.6}$ & (5.1$^{+2.8}_{-1.8}$)$\times10^{8}$ & 300$\pm$40 \\
\multicolumn{11}{c}{{\sc Sample Rejected as Likely Low-$z$ Contaminants}} \\
\galz\ & 09:59:30.49 & $+$02:14:44.10 & 12.38$^{+0.11}_{-0.10}$ & 12.54$^{+0.04}_{-0.07}$ & 12.63$^{+0.10}_{-0.15}$ & $-$21.90$^{+0.15}_{-0.15}$ & $-$2.46$^{+0.22}_{-0.16}$ & 9.0$^{+3.4}_{-4.9}$ & (9.3$^{+0.4}_{-0.5}$)$\times10^{8}$ & 500$\pm$40 \\
\galk\ & 09:59:31.30 & $+$02:08:33.85 & 13.0$^{+1.0}_{-1.0}$ & 12.5$^{+3.5}_{-0.2}$ & 13.8$^{+1.3}_{-1.2}$ & $-$20.97$^{+0.18}_{-0.16}$ & $-$2.44$^{+0.21}_{-0.13}$ & 4.1$^{+1.9}_{-2.1}$ & (4.3$^{+3.7}_{-1.5}$)$\times10^{8}$ & 270$\pm$40 \\
\gali\ & 10:00:20.38 & $+$01:49:58.33 & 14.4$^{+1.5}_{-1.6}$ & 16.7$^{+1.7}_{-3.2}$ & 14.7$^{+3.1}_{-1.6}$ & $-$21.32$^{+0.17}_{-0.22}$ & $-$2.24$^{+0.16}_{-0.20}$ & 11.0$^{+3.2}_{-4.7}$ & (1.1$^{+0.6}_{-0.3}$)$\times10^{9}$ & 260$\pm$70 \\
\enddata
\tablecomments{Positions are measured from the detection image used
  for {\tt SE++} and {\tt SE} classic catalogs.  We provide three
  photometric redshifts for each source from {\tt Bagpipes}, {\tt
    EAzY}, and {\tt LePhare}, but note most of the derived properties
  --- including \muv, $\beta_{\rm UV}$, SFR$_{\rm 100Myr}$, and
  M$_\star$ --- are measured from the best-fit {\tt Bagpipes}
  posterior distributions.  R$_{\rm eff}$ is measured from F277W
  imaging using {\tt Galfit}. We catagorize subsets of our sample as
  described in the text \S~\ref{sec:details}, and include three
  sources which were removed for futher analysis on suspicion they are
  low-$z$ contaminants; if confirmed as high-$z$, their properties may
  reflect what is given in this table.}
\label{tab:phys}
\end{deluxetable*}

%% file: tab_uvlf.tex
\begin{deluxetable}{ccccc}
  \tablecolumns{5}
  \tablecaption{UV Luminosity Function Constraints}
  \tablehead{
\colhead{$z$} & \colhead{$z$ {\sc range}} &     \colhead{$M_{UV}$} & \colhead{$\Delta M_{\rm UV}$} & \colhead{$\Phi$} \\
& & & & [Mpc$^{-3}$\,mag$^{-1}$]\\
  }
  \startdata
\hline
11 & [9.5,\,12.5] & $-$22.0 & 0.8 & (1.0$^{+0.3}_{-0.4}$)$\times10^{-6}$\\
11 & [9.5,\,12.5] & $-$21.2 & 0.8 & (1.4$\pm$0.5)$\times10^{-6}$ \\
\hline
14 & [13,\,15] & $-$21.0 & 1.0 & (8.1$\pm$4.2)$\times10^{-7}$ \\
\enddata
\tablecomments{The measured contribution of our candidates to the UV
  luminosity function at $z=11$ and $z=14$ have not been
  corrected for incompleteness.}
\label{tab:uvlf}
\end{deluxetable}